\documentclass{aa}
\usepackage{amsmath}
\usepackage{amssymb}
\usepackage{xspace}
\usepackage{ulem}
\usepackage{amsmath}
\usepackage{natbib}
\usepackage{array}
\newcommand{\data}{\mathcal{D}}
\newcommand{\Mkpc}{\ensuremath{\text{M}_{1\text{kpc}}}\xspace}

\newcommand{\Lsun}{\ensuremath{\text{L}_{\odot}}}

\usepackage{soul}

\begin{document}

\title{Model comparison of the dark matter profiles of Fornax, Sculptor, Carina and Sextans}
\titlerunning{Model comparison of dark matter profiles in local dSphs}

\author{M. A. Breddels
  \and
  A. Helmi
}
\institute{Kapteyn Astronomical Institute, University of Groningen, P.O. Box 800, 9700 AV Groningen, The Netherlands}
\authorrunning{M.A. Breddels \& A. Helmi}

\abstract{We use orbit based dynamical models to fit the 2nd and 4th moments of
the line of sight velocity distributions of the Fornax, Sculptor,
Carina and Sextans dwarf spheroidal galaxies.  Our goal is to compare
dark matter profile models of these four systems using Bayesian
evidence. We consider NFW, Einasto and several cored profiles for
their dark halos and present the probability distribution functions of
the model parameters.  When considering each system separately, we
find there is no preference for one of these specific parametric
density profiles. However, the combined evidence shows that it is
unlikely that all galaxies are embedded in the same type of cored
profiles of the form $\rho_{DM} \propto 1/(1 + r^2)^{\beta/2}$, where
$\beta = 3, 4$.  For each galaxy, we also obtain an almost model
independent, and therefore accurate, measurement of the logarithmic
slope of the dark matter density distribution at a radius $\sim
r_{-3}$, i.e.  where the logarithmic slope of the stellar density
profile is $ -3$. This slope ranges from $\sim -1.4$ for Fornax to
$\sim -1.1$ for Sextans, both at $\sim 1$ kpc. All our best fit models
essentially have the same mass distribution over a large range in
radius (from just below $r_{-3}$ to the last measured data
point). This remarkable finding likely implies much stronger
constraints on the characteristics that subhalos extracted from
cosmological simulations should have in order to host the dSph
galaxies around the Milky Way.}

\keywords{galaxies: dwarf -- galaxies: kinematics and dynamics}
\maketitle

\section{Introduction}

According to galaxy formation theories dwarf spheroidal galaxies are
believed to inhabit massive dark matter halos. Because of their large
mass to light ratio these galaxies are ideal to test fundamental
predictions of the $\Lambda$CDM cosmological paradigm, since it is generally
considered relatively safe to neglect baryons in the construction of
dynamical models.

One of the strongest predictions from $\Lambda$CDM concerns the
dark matter density profile. Early simulations of dark matter halos
assembled in a cosmological context showed that such a profile is
accurately described by a two-sloped form, now known as NFW profile
\citep{nfw1996, nfw1997}. More recently Einasto profiles have been
shown to provide a better fit \citep[e.g.][]{Springel2008,Navarro2010MNRAS402}, in
particular for satellite galaxies \citep{VeraCiro2013MNRAS}. These
predictions are made using dark matter only simulations and therefore
neglect (by construction) the baryonic component. And although baryons
are sub-dominant in the total potential of the system
\citep{Walker2012}, it has been suggested that they could play a role in the
evolution of dwarf spheroidal galaxies, for instance, in modifying the
internal orbital structure \citep{Bryan2012} and the overall density
profile \citep{Governato2012}. The complex evolution of baryons and
its non-trivial interplay with the host halo are difficult to model
and not yet completely understood \citep[see][]{Ponzen2012MNRAS}.

Another effect driving the internal dynamics of satellite halos is the
tidal interaction with the main host. It can change the density
profile \citep{Hayashi2003}, the geometrical shape of the mass
distribution \citep{Kuhlen2007}, and also influence the kinematics of
the embedded stars \citep{Lokas2010ApJ}. Unfortunately these uncertainties imply that even
when the observations of the local dwarf spheroidal galaxies are not
consistent with being embedded in the halos predicted from pure dark
matter N-body simulations, this does not necessarily reflect a
fundamental problem of $\Lambda$CDM.

Thanks to their relative proximity, information for individual stars
in the dwarf galaxies satellites of the Milky Way are relatively easy
to get. Sky positions are easily determined from photometry, and
radial velocity measurements are possible to estimate within an error
of $\sim 2$ km/s. Some of the
datasets compiled to the date include thousands of individual members
with line-of-sight velocities
\citep{Helmi2006ApJ,Battaglia2006AA,Battaglia2008ApJ...681L..13B,Walker2009AJ....137.3100W,Battaglia2011MNRAS}
Proper motions of individual stars are currently
still too difficult to measure. Despite the fact that only three of
the total of six phase space coordinates are available from
measurements, it is possible to create dynamical models of these
systems that can be compared to these observables.

Following the method thoroughly described in \citet{Breddels2012arXiv}
we set out to model Fornax, Sculptor, Carina and Sextans with
orbit-based dynamical methods (Schwarzschild modeling) assuming they
are embedded in spherical halos. As extensively shown in the
literature
\citep[e.g.][]{Richstone1984ApJ...286...27R,Rix1997ApJ...488..702R,vanderMarel1998ApJ...493..613V,Cretton1999ApJS..124..383C,Valluri2004ApJ...602...66V,vdBosch2008MNRAS.385..647V,Jardel2012}
this method allows to construct a non-parametric estimator of the
distribution function. Among many, this method has one advantage over
Jeans modeling, by not having to assume a particular velocity
anisotropy profile, therefore being more general and thus less prone
to biases associated to the assumptions. But even in this case there
are other limitations in the modeling such as the mass-anisotropy
degeneracy. In this work we use higher moments (fourth moment) of the
line of sight velocity distribution to get a better handle on this
degeneracy.

To compare how different shapes for the dark matter profiles fit the
data, we first need to establish a statistical framework. In this
paper we do this in a Bayesian way using the evidence
\citep{Mackay_2003_information}. This method provides a natural way of
comparing models in Bayesian inference and also makes it possible to
combine the data of all the dwarf spheroidals to test for example, if
all dwarf spheroidals could be embedded in a universal halo
\citep{Mateo1993,Walker2009}. Furthermore, the shape may give us hints
to how the dwarf galaxy formed and the anisotropy profile may be used
to distinguish between evolutionary scenarios \citep[see
e.g.][]{Mayer2010AdAst2010E..25M,Kazantzidis2011ApJ,Helmi2012ApJ}.

This paper is organized as follows. We begin in \S \ref{sec:data} by
presenting the data and all the ingredients needed to do the model
comparison. In \S \ref{sec:method} we present our dynamical and
statistical methods.  We present the results of our Schwarzschild
models for the four dSph in our sample in \S \ref{sec:results_schw},
while the Bayesian model comparison is done in \S \ref{sec:bayes}. We
discuss the implications of our results in \S \ref{sec:slope} and conclude in \S
\ref{sec:end}.

\section{Data}

\label{sec:data}

\begin{figure*}
\centerline{\includegraphics[scale=0.65]{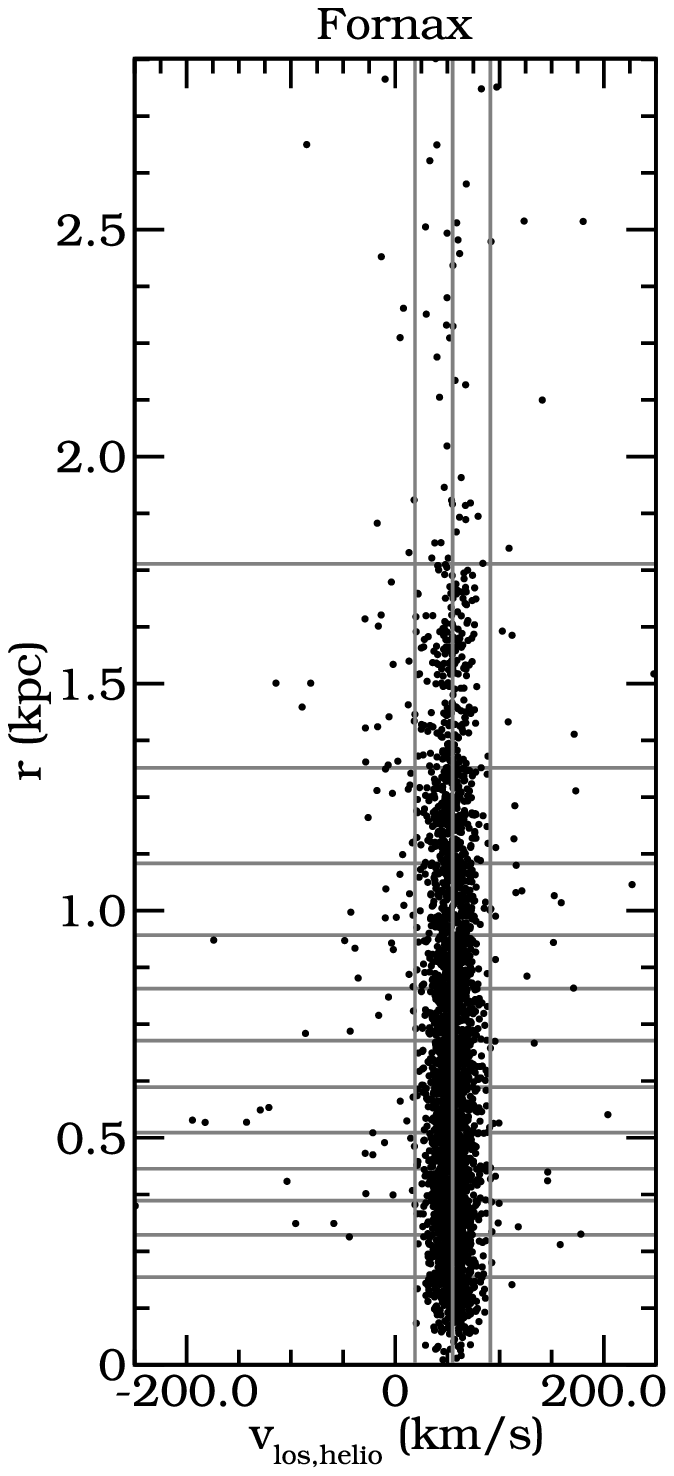}\includegraphics[scale=0.65]{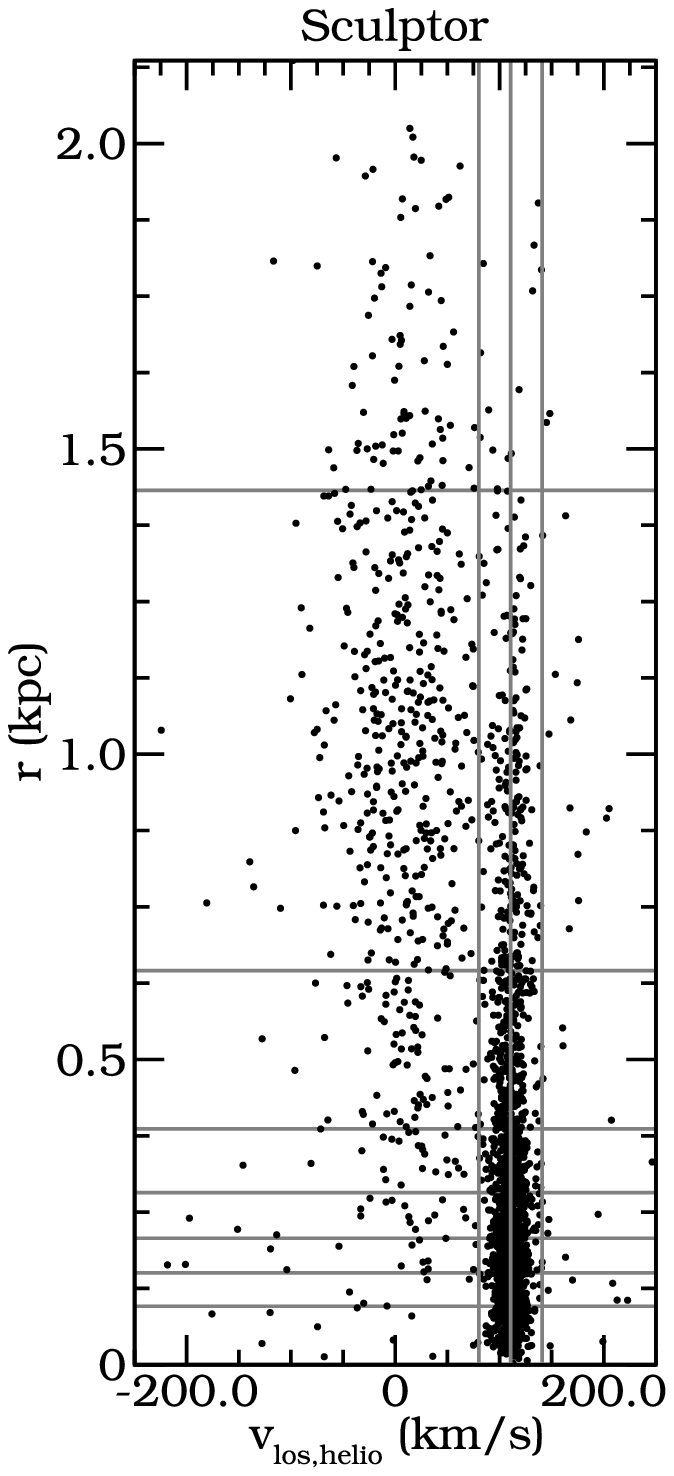}
\includegraphics[scale=0.65]{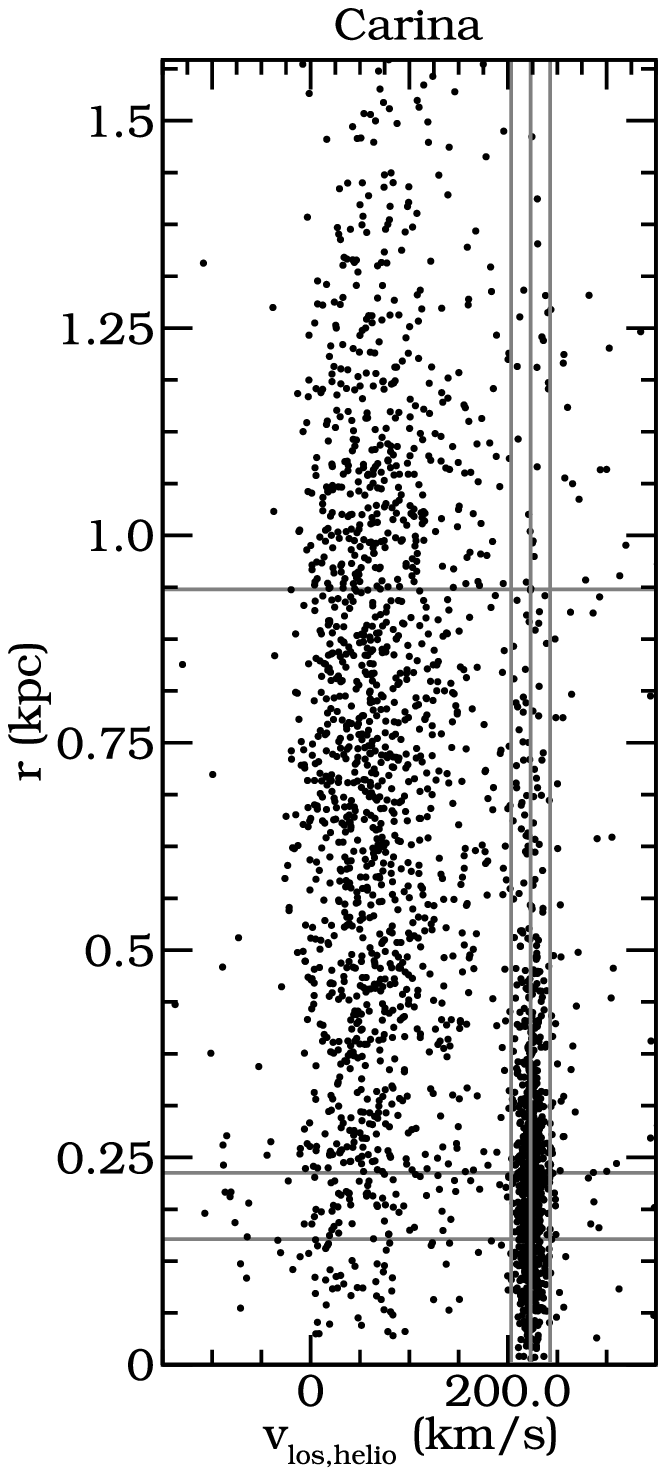}\includegraphics[scale=0.65]{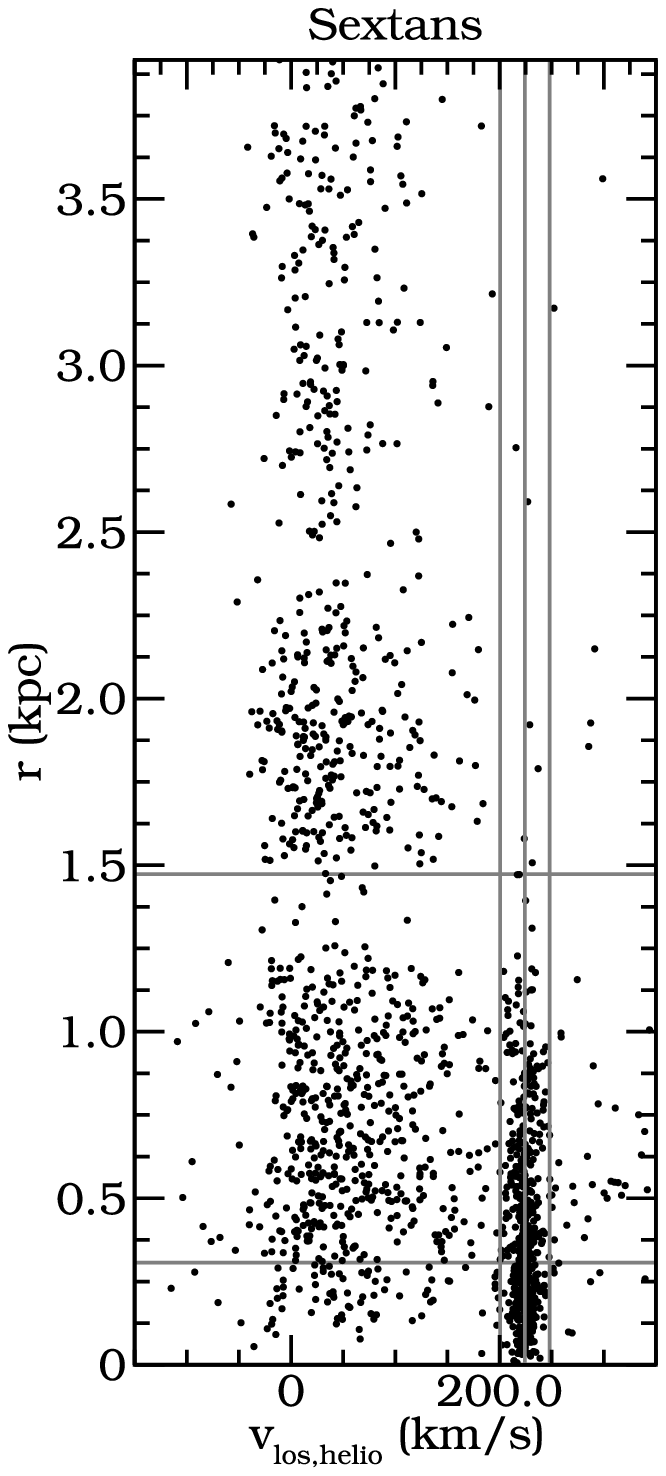}}
\caption{Radius versus line of sight velocity for Fornax, Sculptor Carina and Sextans. The horizontal lines show the borders of the bins, the vertical lines denote the mean systemic velocity of the galaxy together with the $\pm 3\sigma$ region. \label{fig:vlos}}
\end{figure*}

\begin{figure*}
\centerline{\includegraphics[scale=0.45]{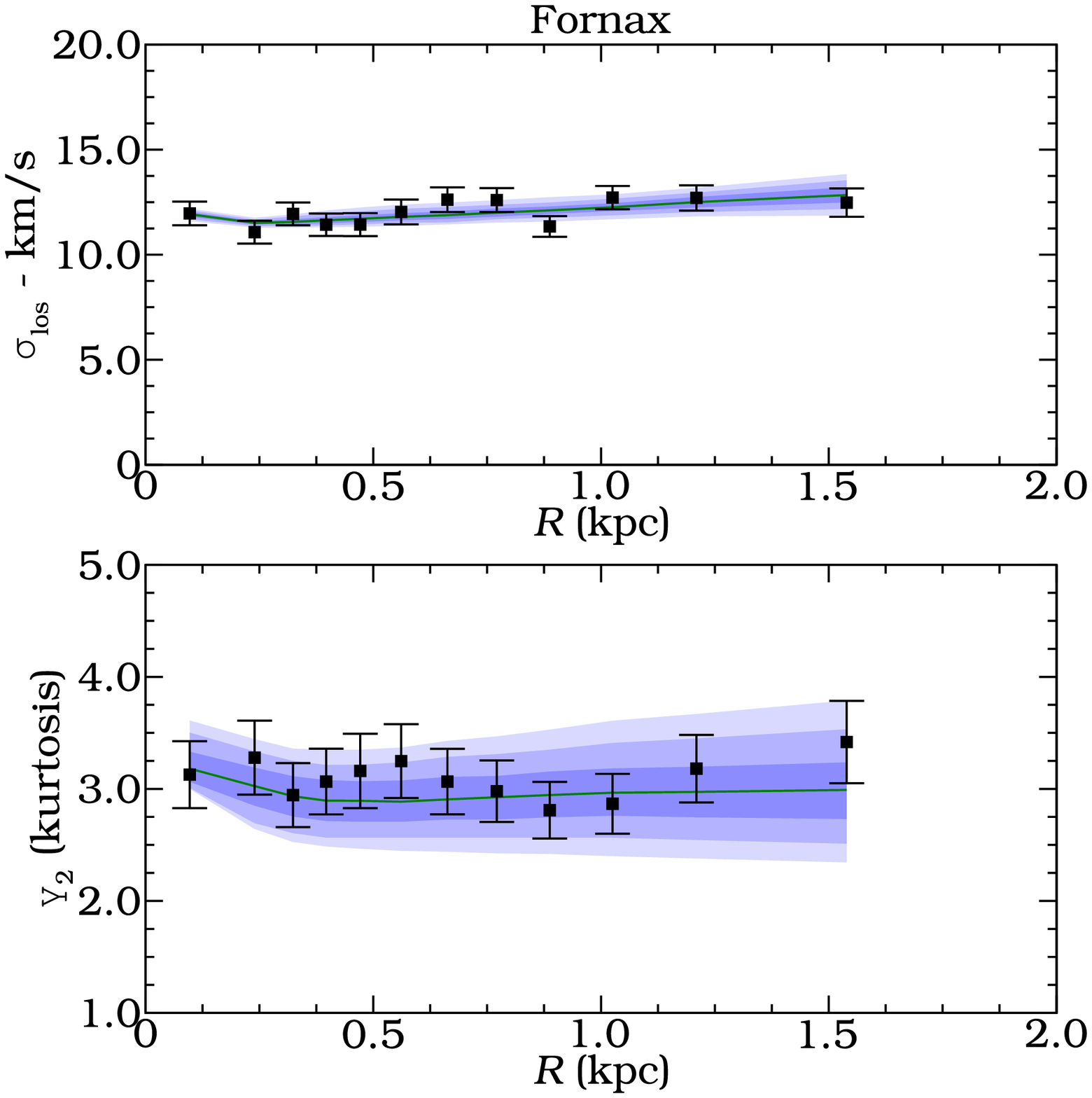}\includegraphics[scale=0.45]{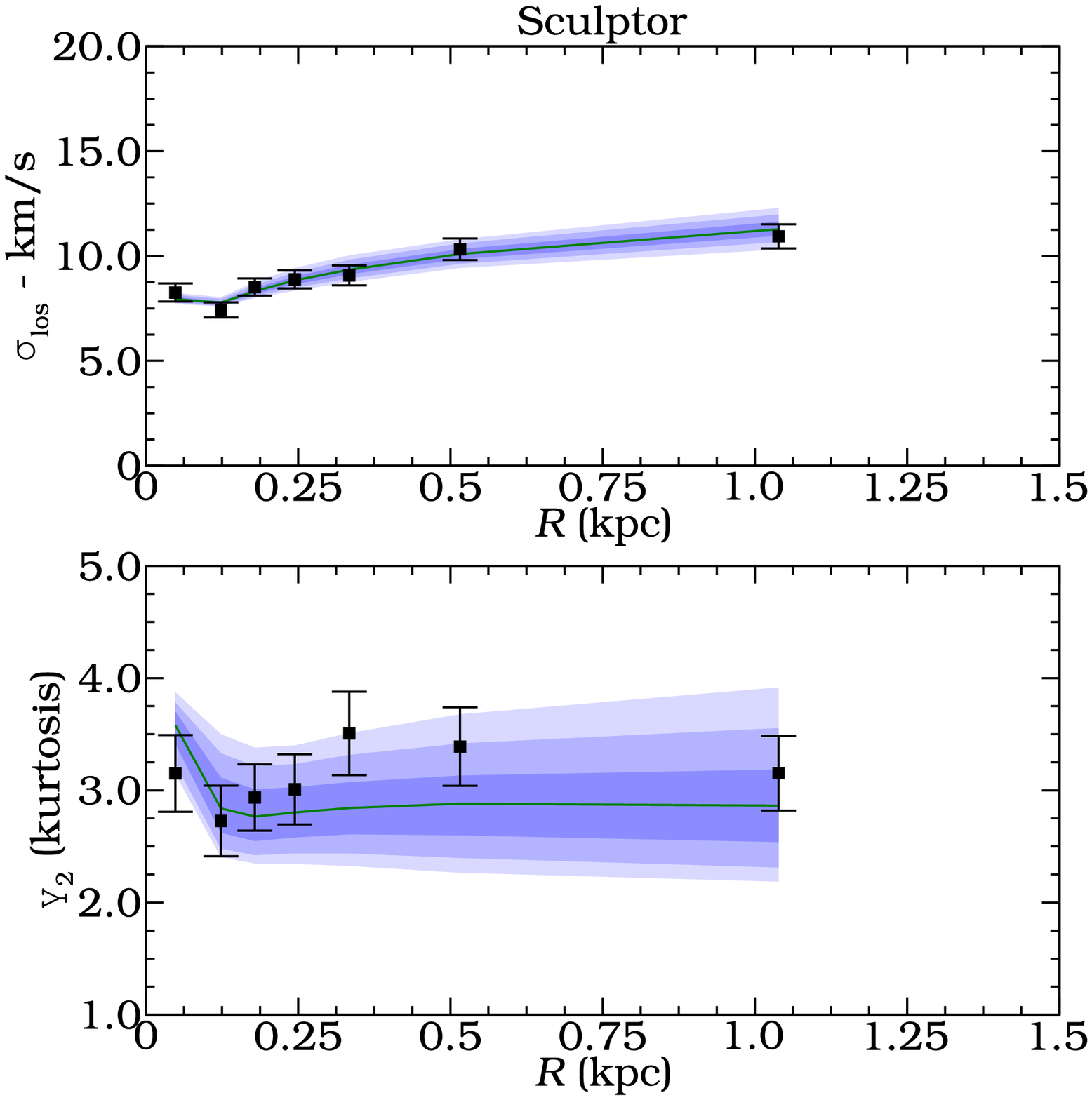}}
\centerline{\includegraphics[scale=0.45]{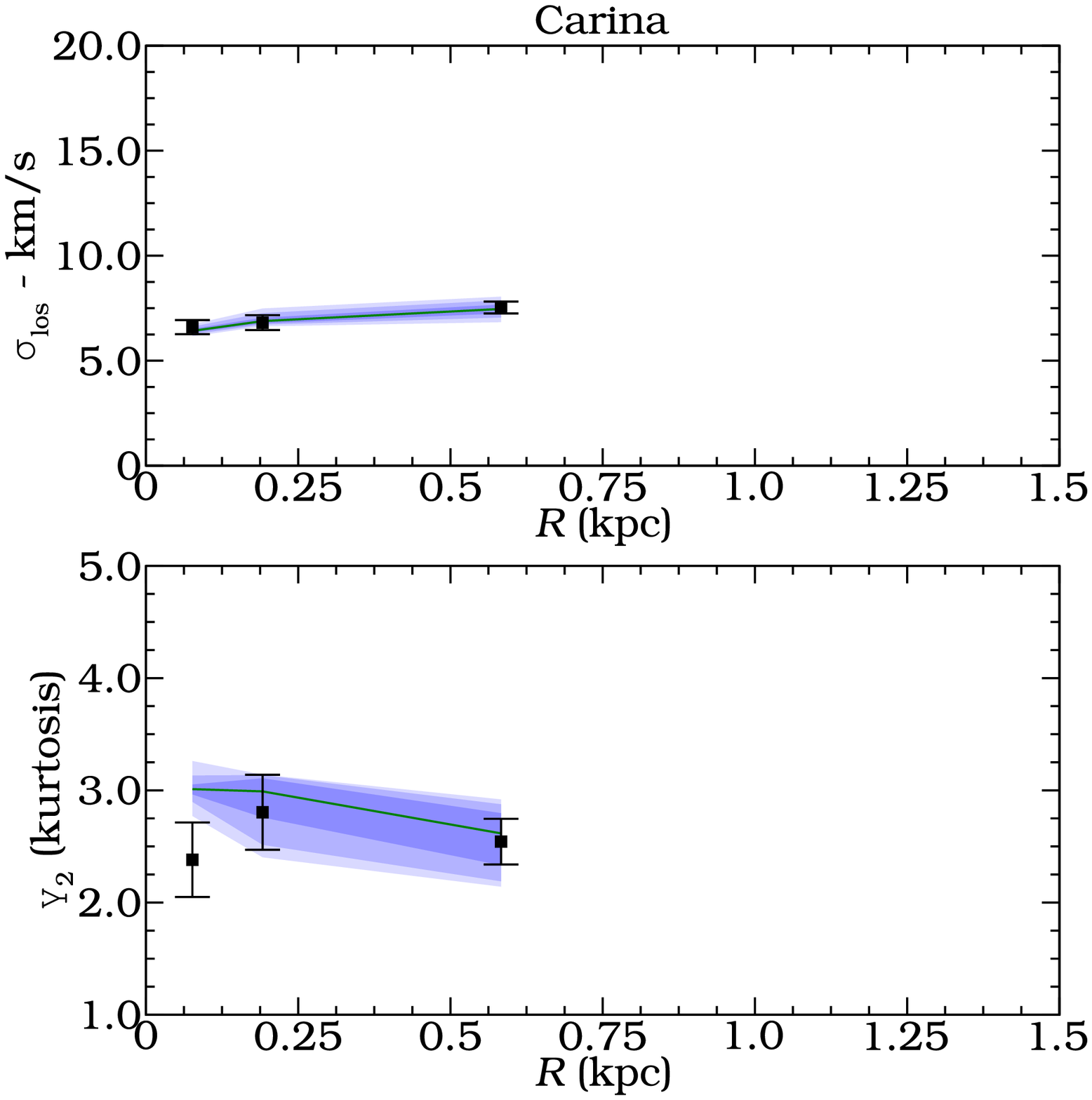}\includegraphics[scale=0.45]{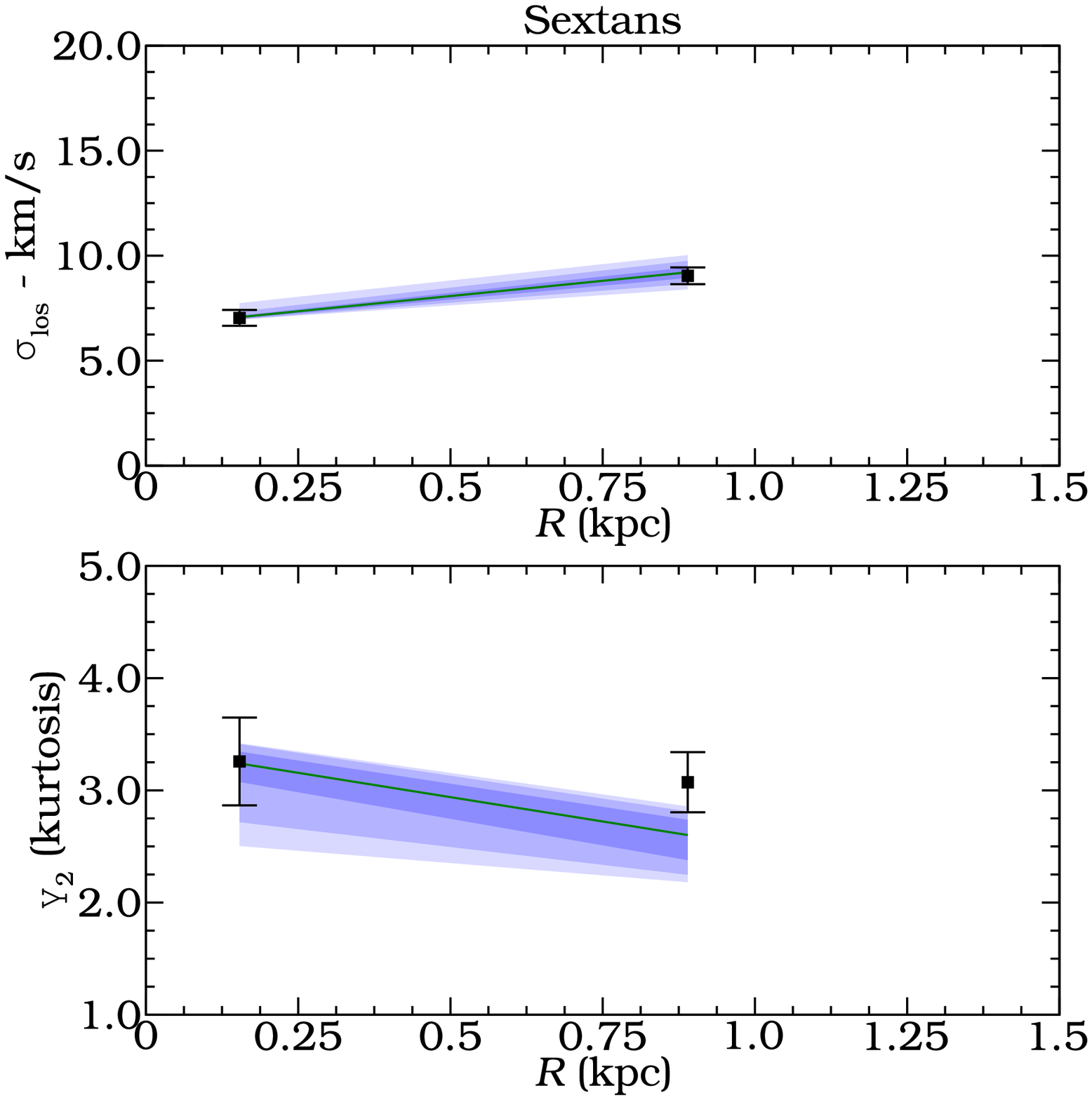}}
\caption{Line of sight velocity moments for Fornax, Sculptor Carina and Sextans. For each galaxy we show the velocity dispersion and the kurtosis. The black dots show the mean, and the error bars the 1$\sigma$ error. The blue regions show the confidence interval for the NFW fit, similar to \cite{Breddels2012arXiv}. \label{fig:moments}}
\end{figure*}

\begin{table}
 \centering
\begin{tabular}{l|r|r|r}
Name & $N_\text{Batt}$ & $N_\text{Walker}$ & $N_\text{member}$ \\
\hline
Fornax 		& 945$^{(1)}$  & 2633$^{(5)}$& 2936 \\
Sculptor	& 1073$^{(2)}$ & 1541$^{(5)}$& 1685 \\
Carina		& 811$^{(3)}$  & 1982$^{(5)}$ & 885 \\
Sextans		& 792$^{(4)}$  & 947$^{(5)}$ & 541 \\
\end{tabular}
\caption{Number of stars in the kinematic samples used in this paper. Sources: $^{(1)}$\citet{Battaglia2006AA}, $^{(2)}$\citet{Battaglia2008ApJ...681L..13B}, $^{(3)}$\citet{Helmi2006ApJ,Koch2006AJ,Starkenburg2010AA}, $^{(4)}$\citet{Battaglia2011MNRAS}, $^{(5)}$\citet{Walker2009AJ....137.3100W}}
\label{tab:data_kin}
\end{table}

\begin{table*}
 \centering
\begin{tabular}{l|c|c|c|c|c|c}
Name & $R_{e,\text{max,Batt}}$ & $R_{e,\text{max,walker}}$ & $\mu_{\rm MW}$  & $\sigma_{\rm MW}$  & $\mu_{\rm dwarf}$ & $\sigma_{\rm dwarf}$ \\
& (kpc) & (kpc) & (km/s) & (km/s) & (km/s) & (km/s) \\
\hline
Fornax & 1.82 & 2.21 & $41.1$  & $ 38.9$   &  $55.1 $  & $12.1$ \\
Sculptor & 1.37 & 1.65 & $17.9$ &$47.4$  & $110.6$  & $ 10.1$ \\
Carina & 0.86 & 0.96 & $70.9$  & $62.5$  &   $222.9$ & $6.6$ \\
Sextans & 1.86 & 1.65 & $67.5$  & $74.5$   & $224.3$ & $7.9$\\
\end{tabular}
\caption{Parameters of the foreground plus dwarf galaxy model used for determining membership, as well as
for deriving the radial profiles for the second and fourth velocity moments for each dSph.}
\label{tab:data_params}
\end{table*}

\begin{table*}
 \centering
\begin{tabular}{l|r|c|c|c}
Name & distance & profile & scale radius & $L_V $ \\
& (kpc) & & (kpc) & $\times 10^5\Lsun$ \\
\hline
Fornax   & 138$^{(1)}$ & Plummer$^2$     & 0.79 & $100^{(2)}$ \\
Sculptor &  79$^{(3)}$ & Plummer$^3$     & 0.30 & $10^{(3)}$ \\
Carina   & 101$^{(1)}$ & Exponential$^4$ & 0.16 & $2.4^{(4)}$\\
Sextans  &  86$^{(1)}$ & Exponential$^4$ & 0.39 & $4.37^{(4)}$ \\
\end{tabular}
\caption{Distances, type of photometric profile used, scale radius and stellar luminosity used for the dynamic models. Sources: $(1)$ \citet{Mateo1998ARA}, $(2)$ \citet{Battaglia2006AA}, $(3)$ \citet{Battaglia2008ApJ...681L..13B}, $(4)$ \citet{Irwin1995MNRAS.277.1354I}}
\label{tab:gal_params}
\end{table*}

In this section we present the data that is used for fitting our
dynamical models. The radial velocity measurements of the dwarf
spheroidal galaxies come from
\citet{Helmi2006ApJ,Battaglia2006AA,Battaglia2008ApJ...681L..13B,Walker2009AJ....137.3100W}
and \citet{Battaglia2011MNRAS}. We plot radius versus heliocentric
velocity in Fig.~\ref{fig:vlos} for each galaxy separately.

Figure \ref{fig:vlos} shows that each dSph suffers from foreground
(Milky Way) contamination. To remove this contamination and reliably
identify member stars we have developed a simple analytic model for
the positional and kinematic distribution of both foreground and the
galaxy in question \citep[along the lines of][Breddels et al.\ in
prep]{Battaglia2008ApJ...681L..13B}.  For each particular
dataset\footnote{For a given dSph there may be multiple datasets, and
  we treat each independently because their sampling might be
  different.}, we assume that the foreground has a constant surface
density, and that the dSph follows a specific stellar density
profile. We also assume that the velocity distribution at each radius
may be modeled as sum of two Gaussians. The Gaussian describing the
foreground has the same shape at all radii, while that of the stars
associated with the dwarf can have a varying dispersion with
radius. Their relative amplitude also changes as function of
distance from the dwarf's centre. This model results in a
determination of the relative contribution of member-to-non-member
stars as a function of velocity and radial distance $R$.

Based on this model we calculate the elliptical radius at which the
ratio of dSph:foreground is 3:1 (without using any velocity
information). We remove all stars outside of this radius from the
dataset. A particular star included in more than one dataset is
removed only when it it satisfies the condition for all sets, for
instance a star outside $R_{e,\text{cut, Batt}}$, but inside
$R_{e,\text{cut, Walker}}$ will not be discarded. This simple clipping
in elliptical radius cleans up part of the foreground
contamination. For completeness, the radii for all datasets cleaned up
in this way are presented in Table \ref{tab:data_params}, as well as
the fit to the foreground model. The number of stars and the sources
are listed in Table \ref{tab:data_kin}.

For the resulting dataset, we compute the second and fourth moment of
the radial velocity as a function of circular radius as
follows\footnote{ Elliptical radii are only used for the clipping, for
  the rest of the analysis we use the circular radius}.  We first
define radial bins such that each bin has at least 250 stars in the
velocity range $v_\text{sys}-3\sigma_v,v_\text{sys}+3\sigma_v$. If the
last bin has less that 150 objects, the last two bins are
merged. After this, we fit our parametric model for the galaxy plus
foreground for each radial bin, to derive new velocity
dispersions. Then for each bin we do a $3\sigma$ clipping on the
velocity using the new velocity dispersion, and from this selection we
calculate the second and fourth moments. The errors on the moments are
computed using Eqs.~(17) and (19) in
\citet{Breddels2012arXiv}. The second moment and the kurtosis\footnote{The kurtosis is defined as $\gamma_2 = \mu_4/\mu_2^2$, where $\mu_4$ is the fourth and $\mu_2$ is the second moment of the line of sight velocity distribution.}  are
shown in Fig. \ref{fig:moments} for each galaxy, where the black dot
corresponds to the mean, and the error bars indicate the 1$\sigma$
error bar. The blue region shows the confidence interval for the NFW
model found in \S \ref{sec:results_schw}.

For the photometry we use analytic fits given by various literature
sources as listed in Table \ref{tab:gal_params}. Although the stellar
mass is sub-dominant in the potential, we do include its contribution
in the dynamic models and fix $M/L_V=1$, as in
\citet{Breddels2012arXiv}.

\section{Methods}
\label{sec:method}

\subsection{Dynamical models}

Our aim is to compare different models to establish what type of dark
matter profile best matches the kinematical data of local dSph
galaxies.  Here we consider the following profiles to describe the
dark matter halos of the dwarfs in our sample:
\begin{align}
\rho(r) &= \frac{\rho_0}{ x \left(1+x \right)^2}, && \text{NFW}\\
\rho(r) &= \frac{\rho_0}{\left(1+ x^\gamma \right)^{\beta/\gamma}}, && \text{(cored) $\beta\gamma$-profile} \\
\rho(r) &= \rho_0 \exp \left( -\frac{2}{\alpha'} \left( x^{\alpha'} -1 \right) \right), && \text{Einasto}
\end{align}
where $x=r/r_s$ and $r_s$ is the scale radius. Each model has at least
two unknown parameters $r_s$ and $\rho_0$. As we did in
\citet{Breddels2012arXiv}, we transform these two parameters to $r_s$
and \Mkpc (the mass within 1 kpc).  As discussed in the Introduction,
the NFW and Einasto models are known to fit the halos dark matter distributions extracted from
cosmological N-body simulations. On the other hand, we explore the $\beta\gamma$ models
to test the possibility of a core in the dark halo. Note that, in comparison to the NFW profile, the Einasto
model has one extra parameter ($\alpha'$), but here we consider only
two values for $\alpha'=0.2, 0.4$ to cover the range suggested by
\citet{VeraCiro2013MNRAS}. On the other hand, the $\beta\gamma$
profiles have two extra parameters, but we limit ourselves here to two
different outer slopes ($\beta=3,4$) and two different transition
speeds between the inner and the outer slopes ($\gamma=1,2$). Note that the $\beta\gamma$
models have a true core only for $\gamma > 1$, however in all cases the central logarithmic slope vanishes,
$d \log \rho/d\log r = 0$. However, we loosely refer to these models as cored in what follows.
Note that, with these choices,  all of our profiles
ultimately have just two free parameters.  The list of models explored and their
parameters are summarized in Table
\ref{tab:model_names}.

\begin{table}
 \centering
\begin{tabular}{l|c|l}
Name & Fixed parameters & Free parameters \\
\hline
NFW & - & \Mkpc, $r_s$ \\
core13 & $\beta=3,\gamma=1$ & \Mkpc, $r_s$ \\
core14 & $\beta=4,\gamma=1$ & \Mkpc, $r_s$ \\
core23 & $\beta=3,\gamma=2$ & \Mkpc, $r_s$ \\
core24 & $\beta=4,\gamma=2$ & \Mkpc, $r_s$ \\
einasto.2 & $\alpha'=0.2$ & \Mkpc, $r_s$ \\
einasto.4 & $\alpha'=0.4$ & \Mkpc, $r_s$ \\
\end{tabular}
\caption{Model names and their characteristic parameters of the various dark matter density profiles explored.}
\label{tab:model_names}
\end{table}

The orbit-based dynamical (Schwarzschild) models of each dwarf galaxy
are obtained as follows \citep[see][for a more detailed
description]{Breddels2012arXiv}.  For each of the dark halo profiles,
with its own set of parameters, we integrate a large number of orbits
in the respective gravitational potential (including also the
contribution of the stars).  We then find a linear combination of
these orbits that fits both the light and the kinematics. The orbital
weights found in this way have a physical meaning and can be used to
obtain the distribution function of the system. As data we have the
line of sight velocity moments (second and fourth depicted in
Fig. \ref{fig:moments}), and the light profile (Table
\ref{tab:data_params}).  The best fit models (which give us the values of the
parameters for a specific dark matter halo profile) are those that
minimize the $\chi^2 = \chi^2_{\rm kin} + \chi^2_{\rm reg}$, under the
condition that the orbital weights are positive, and that the observed
light distribution is fit to better than 1\% at each radius. Here
$\chi^2_{\rm kin} = \sum_k (\mu_{2,k} -
\mu^{model}_{2,k})^2/{\rm var}(\mu_{2,k}) + \sum_k (\mu_{4,k} -
\mu^{model}_{4,k})^2/{\rm var}(\mu_{4,k})$. The $\chi^2_{\rm reg}$ is a
regularization term to make sure that the solution for the orbit
weights leads to a relatively smooth distribution function.
\citet{Breddels2012arXiv} calibrated the amplitude of this term for
Sculptor.  To have the regularization term for the other dwarfs of the
same relative strength, we note that $\chi^2_\text{reg} \propto 1/N$,
where $N$ is the number of members with radial velocities, since the
$\chi_\text{kin}^2$ term also scales as $1/N$. Therefore, normalizing
its amplitude to that of Sculptor we get $\chi^2_\text{reg, dwarf} =
\chi^2_\text{reg, \text{Scl}} \times N_\text{Scl}/N_{\rm dwarf}$.

\begin{figure*}
\begin{tabular}{c @{}c@{} @{}c@{} @{}c@{} @{}c@{}}
\textbf{Fornax} & \textbf{Sculptor} & \textbf{Carina} & \textbf{Sextans}\\
\includegraphics[scale=0.6]{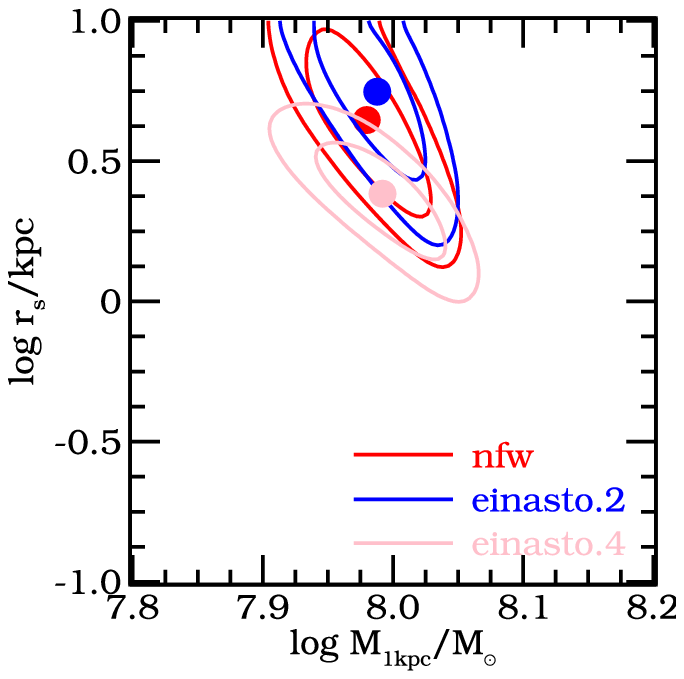} & \includegraphics[scale=0.6]{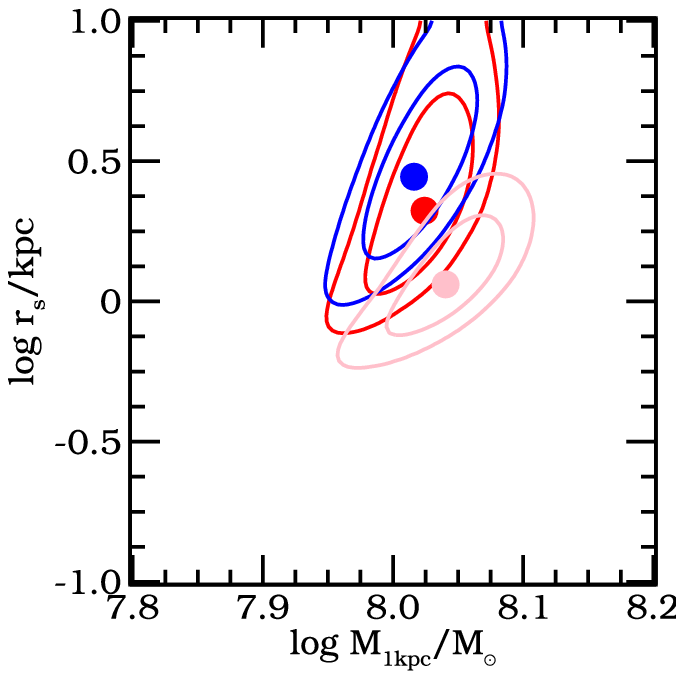} & \includegraphics[scale=0.6]{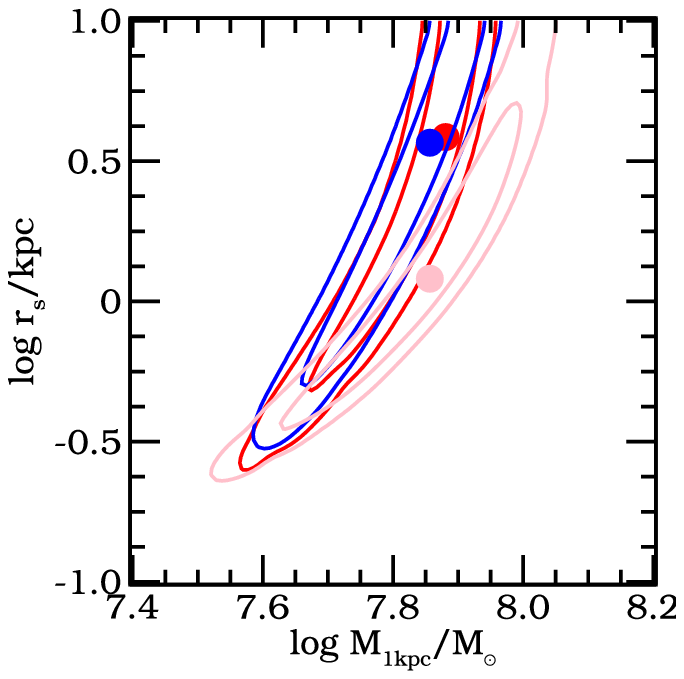} & \includegraphics[scale=0.6]{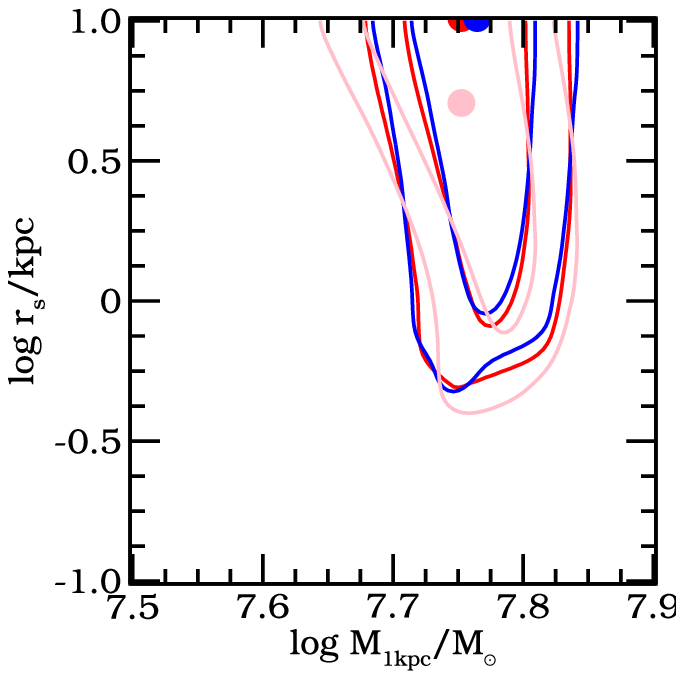} & \rotatebox{90}{\hspace{1cm}\textbf{NFW/Einasto}}\\
\includegraphics[scale=0.6]{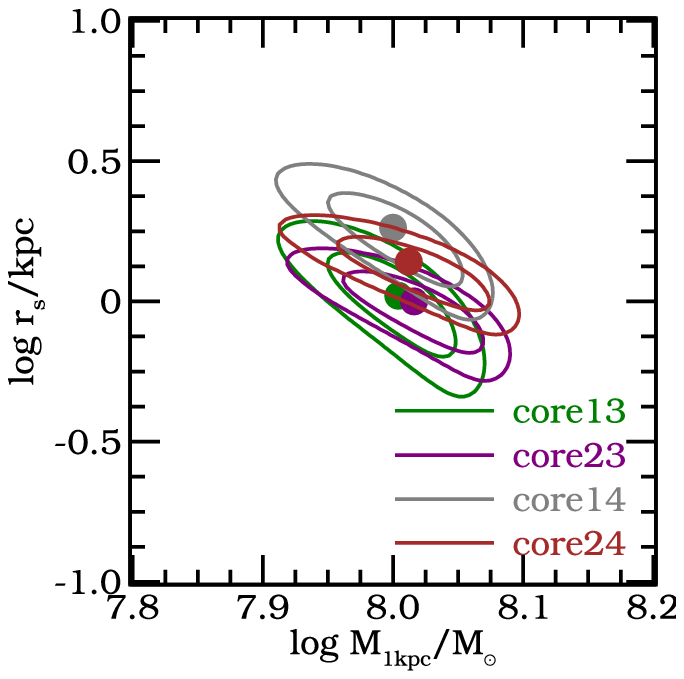} & \includegraphics[scale=0.6]{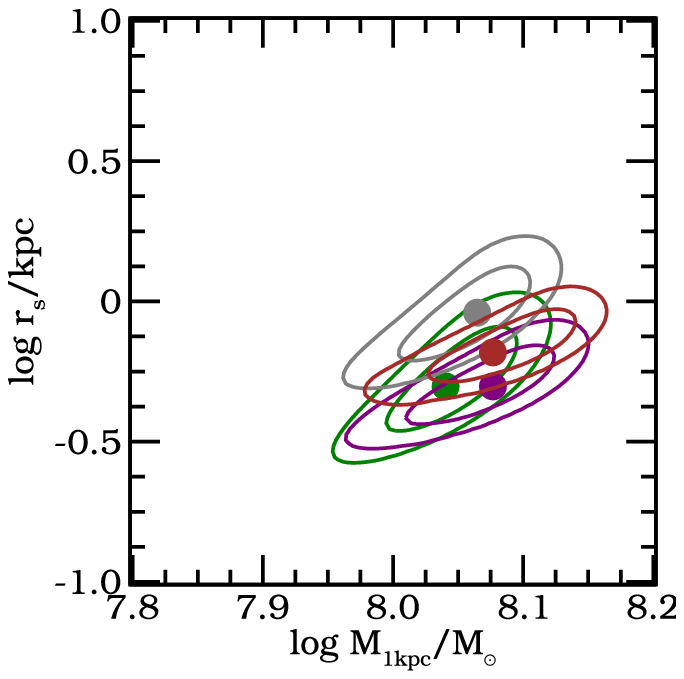} & \includegraphics[scale=0.6]{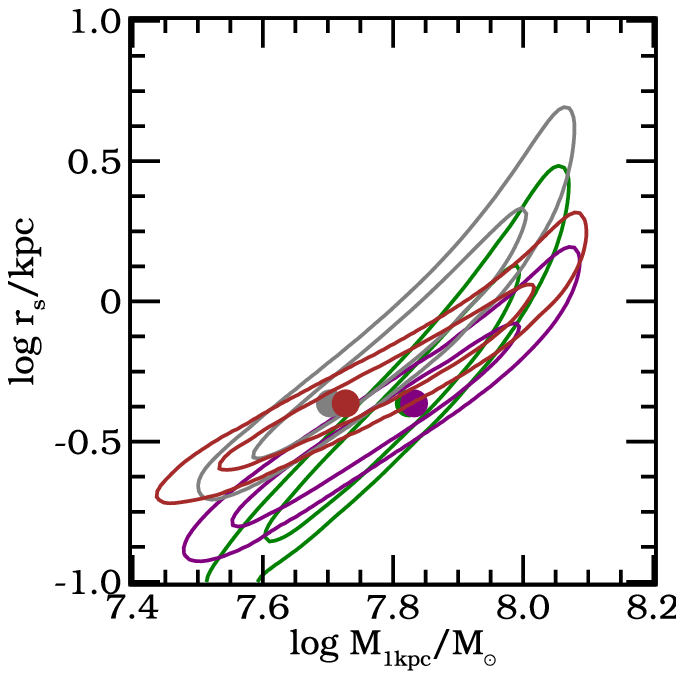} & \includegraphics[scale=0.6]{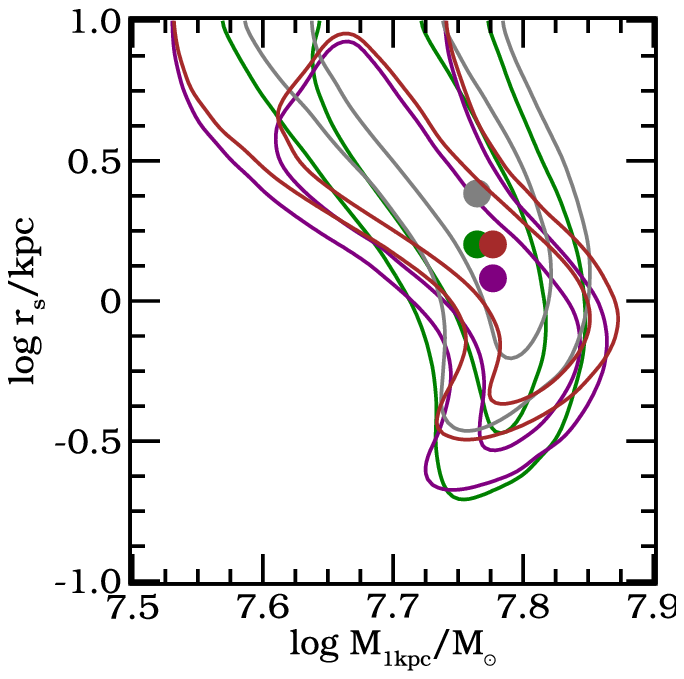} & \rotatebox{90}{\hspace{2cm}\textbf{cored}}\\
\end{tabular}
\caption{Pdf for the two free parameters characterizing the dark halo
  profiles for each dSph galaxy obtained using Schwarzschild
  modeling. The top row shows the pdfs with NFW/Einasto models, the
  bottom panel those for all cored models explored. The contours show
  the 1 and 2$\sigma$ confidence levels (the 3$\sigma$ contour is not
  shown to avoid crowding the image). \label{fig:pdfs}}
\end{figure*}

\subsection{Bayesian model comparison}

Background on Bayesian model comparison may be found in
\citet{Mackay_2003_information}. For completeness we discuss it here
briefly, but we assume the reader is familiar with the basics of Bayesian
inference.

Given the data $\data$ and assuming a model $M_i$, the posterior for the parameters $\theta_i$ of this model is:
\begin{equation}
p(\theta_i|\data, M_i) = \frac{p(\data|\theta_i, M_i) p(\theta_i|M_i)}{p(\data|M_i)}.
\end{equation}
The normalization constant $p(\data|M_i)$, also called the evidence, is of little interest in parameter inference, but is useful in Bayesian model comparison. To assess the probability of a particular model given the data, we find
\begin{equation}
p(M_i|\data) = \frac{p(\data|M_i)p(M_i)}{p(\data)}, 
\end{equation}
where we see the evidence is needed. In this case $p(\data)$ is the uninteresting normalization constant, as it cancels out if we compare two models:
\begin{equation}
\frac{p(M_i|\data)}{p(M_j|\data)} = \frac{p(\data|M_i)}{p(\data|M_j)}\frac{p(M_i)}{p(M_j)} = B_{i,j}\frac{p(M_i)}{p(M_j)}, \label{eq:bayes} 
\end{equation}
where $B_{i,j}$ is called the Bayes factor. If we take the priors on the different models to be equal (i.e. $p(M_i)=p(M_j)$), the ratio of the evidence (the Bayes factor $B_{i,j}$) gives the odds ratio of the two models given the data $\data$.

Using these results we can perform model comparison between dark
matter density profiles, i.e. $\mathcal{M}=\{M_\text{nfw},
M_\text{Einasto}, ...\}$, and calculate for instance the odds that a
given galaxy is embedded in an NFW profile compared to an Einasto
model, $B_{\text{NFW},\text{Einasto}}$.

Not only can we do model comparison on a single object, but we may
also test if our objects share a particular model (e.g. they are all
embedded in NFW halos). If our dataset $\data$ consists of the
observations of two galaxies, i.e. $\data = \data_1 \cup \data_2$ and
assuming the datasets are uncorrelated and independent, we obtain:
\begin{equation}
\frac{p(M_i|\data)}{p(M_j|\data)} = \frac{p(\data_1|M_i)p(\data_2|M_i)}{p(\data_1|M_j)p(\data_2|M_j)}\frac{p(M_i)}{p(M_j)} = B_{i,j,1}B_{i,j,2}\frac{p(M_i)}{p(M_j)} \label{eq:bayes_mul}
\end{equation}
where each factor $p(\data_k|M_i)$ should be marginalized over its (own) characteristic parameters. From Eq.~(\ref{eq:bayes_mul}) we can see that the odds ratio of the models and Bayes factor from different measurements can be multiplied to give combined evidence for a particular model.

Behind each $p(M_i|\data)$ is a set of orbit based dynamical
(Schwarzschild) models, obtained as described above.  For each of the
models we calculate the evidence. Later on we compare each model's evidence to that
of an NFW profile, i.e.\ we compute the Bayes factor $B_{i,\text{NFW}}$,
where $i$ can be e.g.\ Einasto. By definition
$B_{\text{NFW},\text{NFW}}=1$, and again assuming equal priors on the
different models, the Bayes factor equals the odds ratio of the
models, such that for $B_{i,\text{NFW}} > 0$, model $i$ is favored
over an $\text{NFW}$ profile.

\section{Results}
\label{sec:apply}


\subsection{Schwarzschild models}
\label{sec:results_schw}

\begin{figure*}
\centerline{\includegraphics[scale=0.45]{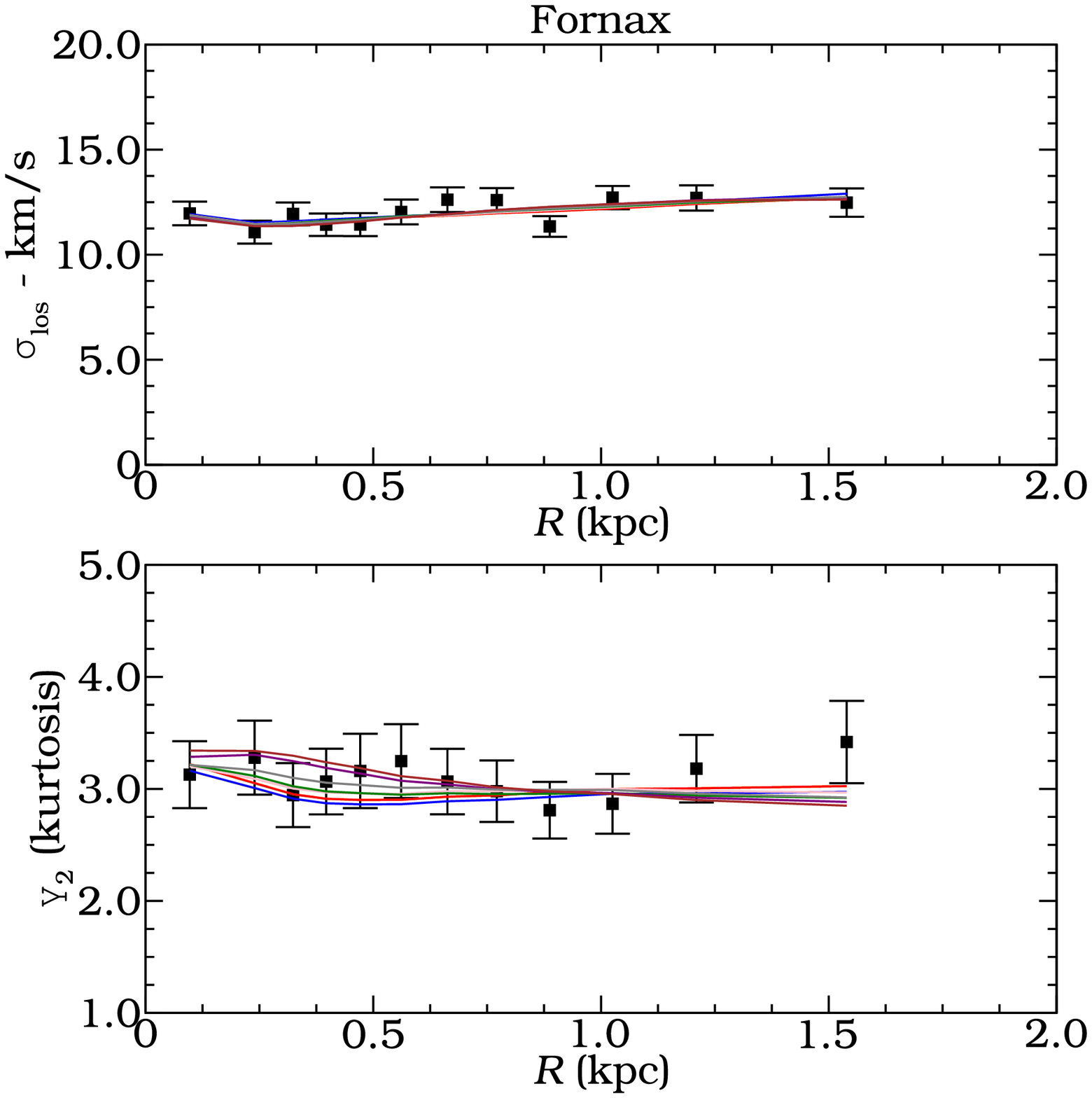}\includegraphics[scale=0.45]{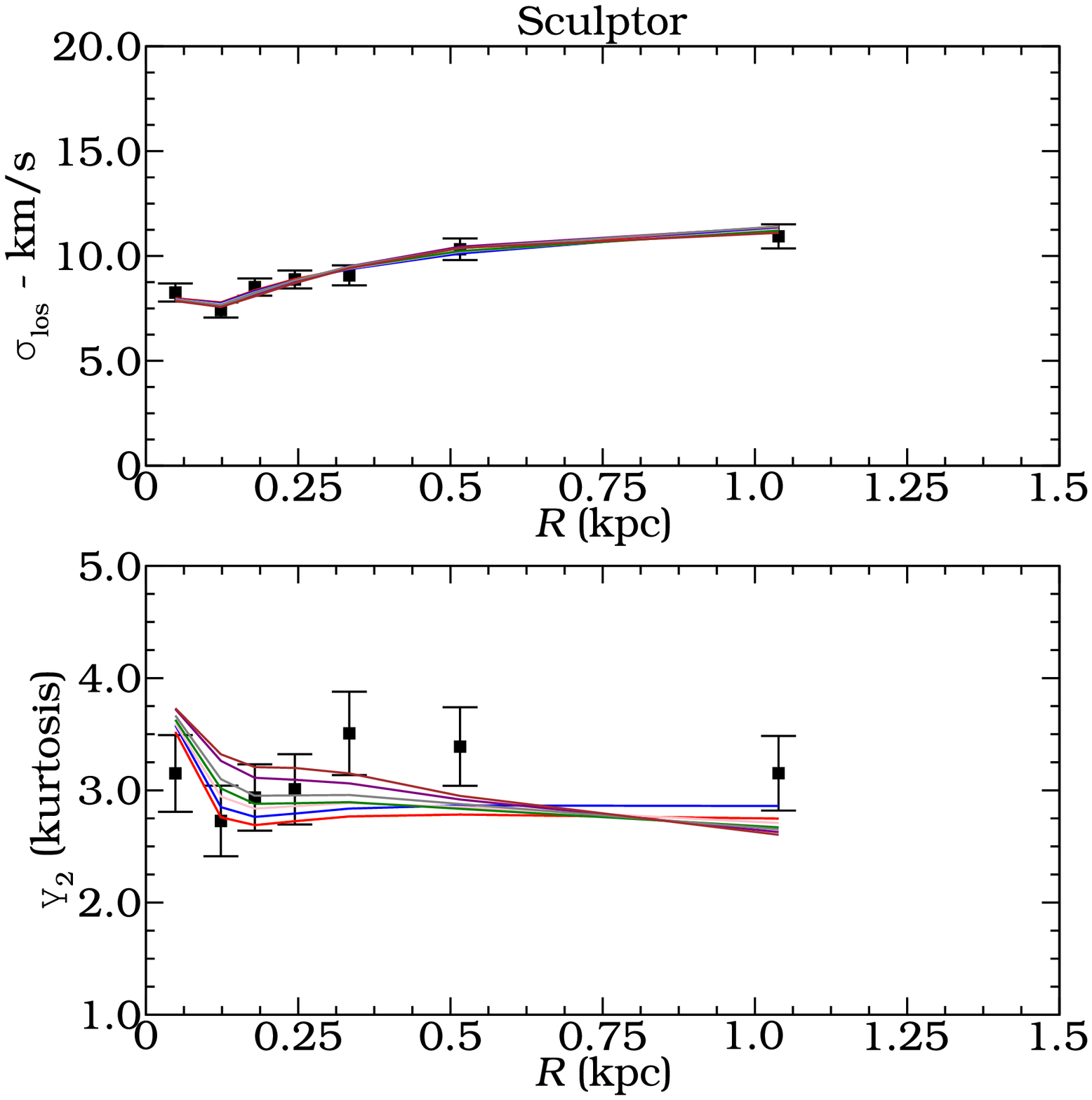}}
\centerline{\includegraphics[scale=0.45]{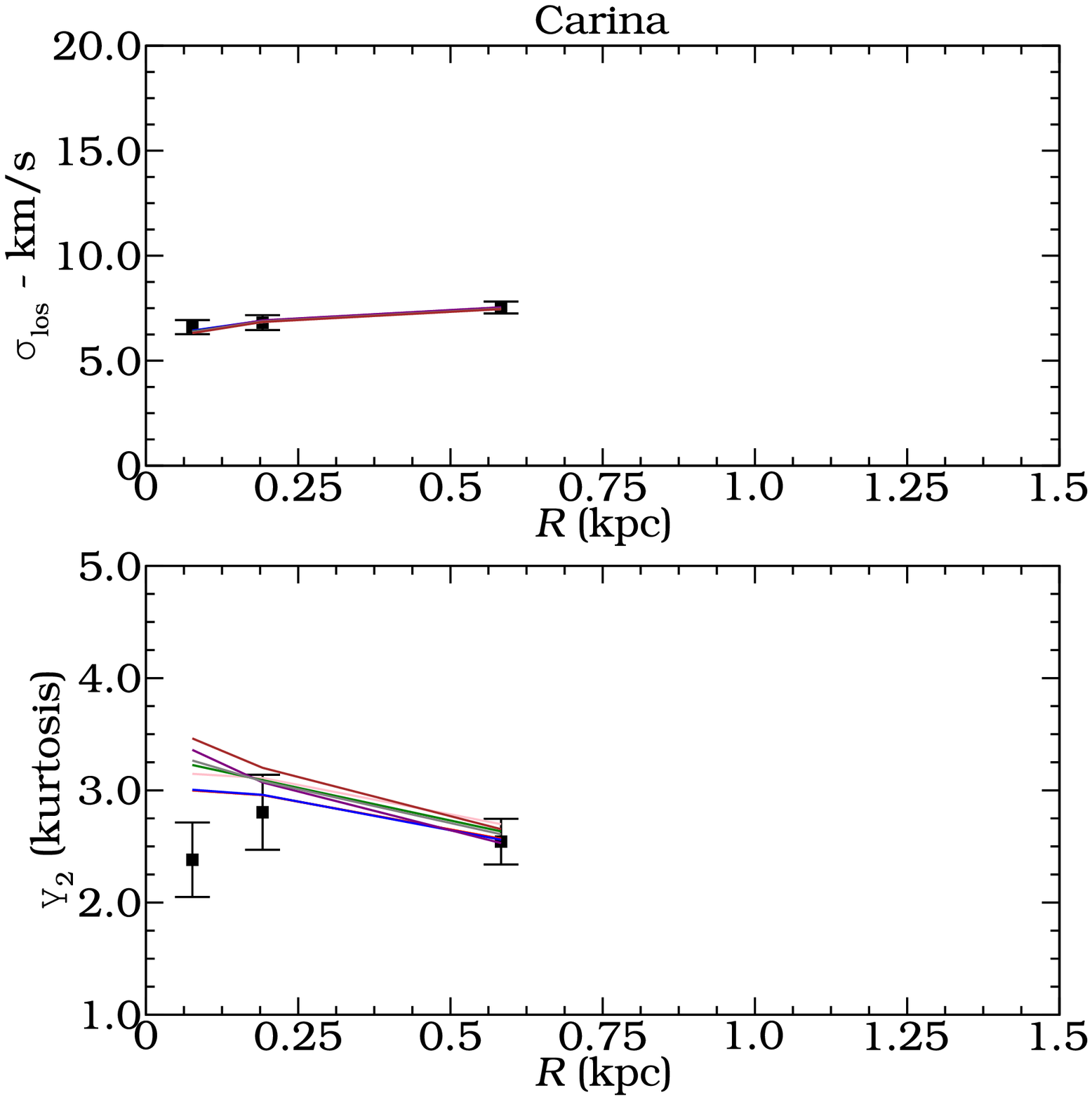}\includegraphics[scale=0.45]{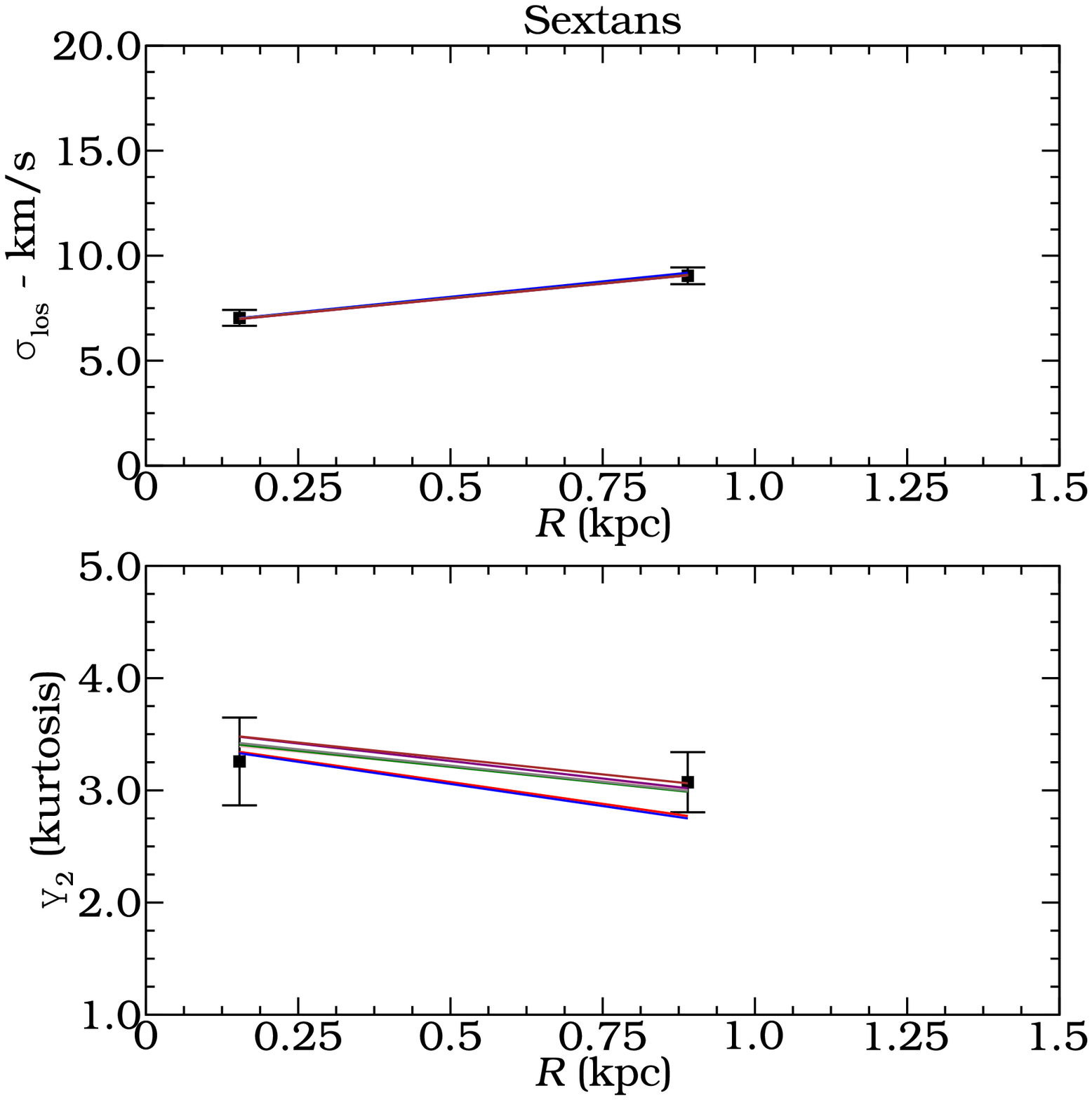}}
\caption{Similar to Fig. 2, except now we show the different best fit models for the various density profiles explored, which are indicated with different colors (the color scheme is the same as in Fig.~\ref{fig:pdfs}). \label{fig:moments_bestfit}}
\end{figure*}

\begin{figure*}
\centerline{\includegraphics[scale=0.6]{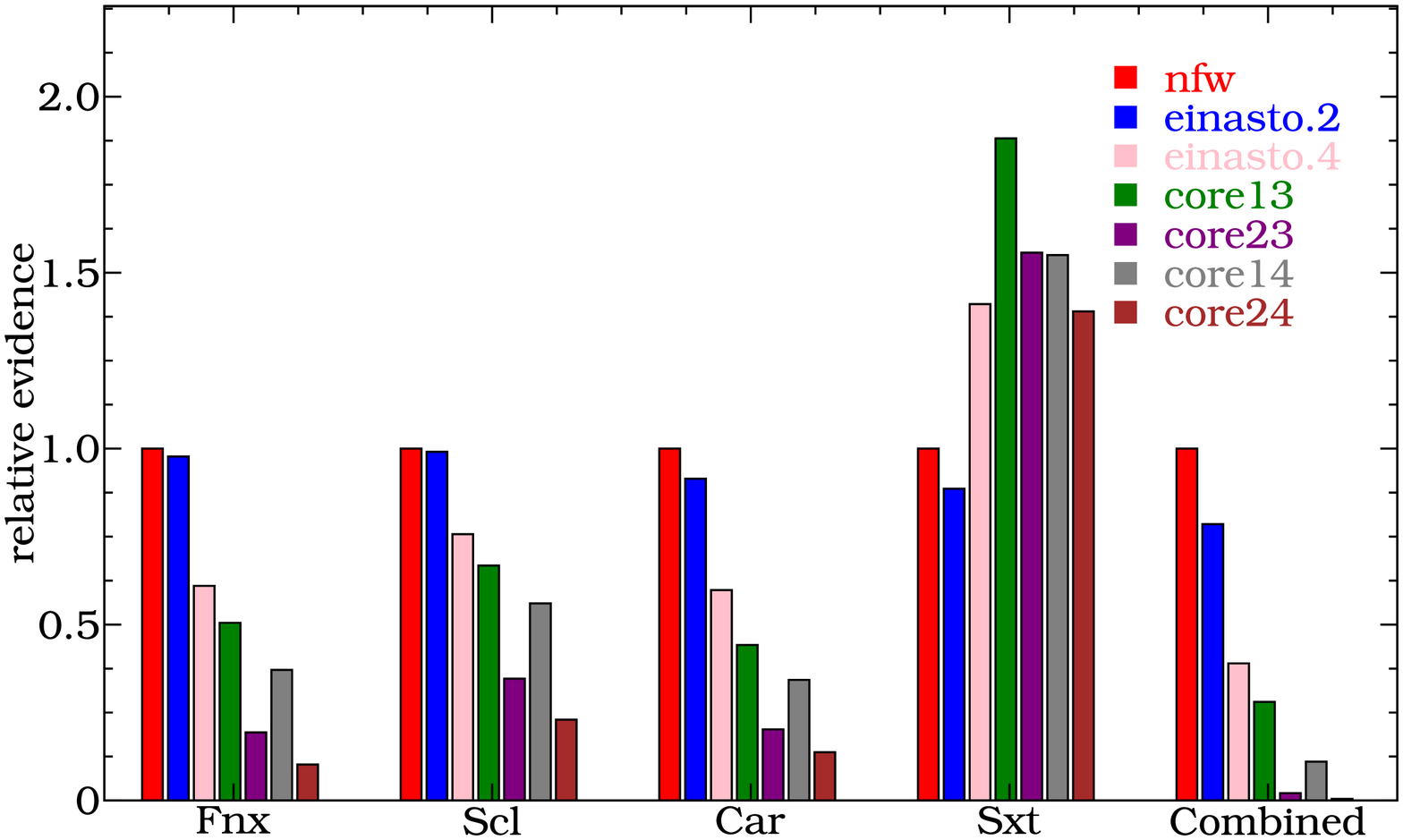}}
\caption{Evidences for all models listed in Table \ref{tab:model_names}, relative to the NFW case. The last column shows the combined evidence for all galaxies together, and shows that the core23 and core24 are strongly disfavored.\label{fig:evidence}}
\end{figure*}

As a result of our Schwarzschild modeling technique, we obtain a two
dimensional probability density function (pdf) of the two parameters,
\Mkpc and $r_s$, for each galaxy and for each dark matter halo profile. In Fig. \ref{fig:pdfs} we plot the
pdf for the cored models and the NFW and Einasto models separately for
each galaxy. The colored dots correspond to the maximum likelihood for
each of the corresponding models as indicated by the legend. The
contours show the 1 and 2$\sigma$ equivalent confidence intervals
(the 3$\sigma$ contour is not shown for clarity). For both Fornax and
Sculptor the parameters for all profiles are relatively well
determined, while for Carina and Sextans this is less so. This can be
attributed to the difference in sample size (and hence to the smaller
number of members) in these systems, which has translated into fewer
bins where the moments can be computed (see
Fig. \ref{fig:moments}). In general for all four galaxies the scale
radius for the cored profiles is found to be smaller than that for the
NFW/Einasto profiles. We come back to this point in section \ref{sec:slope}.

Our model's masses at $r_{1/2}$, the 3d radius enclosing half of the stellar mass,
are compatible with those of \citet{Wolf2010MNRAS}. However, our results for Fornax do not agree
with those of \citet{Jardel2012}. These authors prefer a cored profile with a much
larger scale radius, and their enclosed mass is smaller in comparison to \citet{Wolf2010MNRAS}. We note that
this might be partly related to the fact that the amplitude of their line of sight
velocity dispersion profile  (see their Fig. 2) is slightly lower than what we have determined here.

In Fig.~\ref{fig:moments_bestfit} we overlay on the kinematic observables the predictions
from the best fit Schwarzschild models. We note that all models provide very similar
and virtually indistinguishable fits, especially for the 2nd moment. Some slight differences
are apparent in the kurtosis, but in all cases, the differences are smaller than the error bars
on the moments.

In general we find all anisotropy profiles to be roughly constant with
radius and slightly tangentially biased on average. We do not find
significant differences between the profiles for cored and NFW models (the
reason for this will become clear in \S \ref{sec:slope}).
Fornax' s anisotropy $\beta \sim -0.2\pm0.2$, while Sculptor and Carina
have on average $\beta \sim -0.5\pm0.3$. For Sextans the anisotropy
cannot be determined reliably, $\beta \sim -0.3\pm0.5$. These values are
compatible with those of \citet{Walker2007ApJ}, which were derived using the spherical Jeans
equation assuming a constant velocity anisotropy profile.

\subsection{Bayesian comparison of the models}
\label{sec:bayes}

We compute the evidence relative to the NFW using Eq.~(\ref{eq:bayes})
by integrating over the parameters (in our case the scale radius and
the mass) the pdfs shown in Fig.~\ref{fig:pdfs}. We do this for each
dwarf galaxy and for all the models listed in Table
\ref{tab:model_names}. The different Bayes factors are shown in
Fig.~\ref{fig:evidence}. Each set of bars shows the Bayes factors for
the given dSph galaxy ($B_{i,NFW,k}$), while the last set shows the
combined result ($B_{i,NFW,\text{comb}} = \prod_k B_{i,NFW,k}$). We
note that an odds ratio between 1:2 till 1:3 is considered ``Barely
worth mentioning'' \citep{jeffreys1998theory}, and only odds ratios
above 1:10 are considered ``strong'' evidence.

For each galaxy there is hardly any evidence for or against an Einasto
profile (with $\alpha'=0.2,0.4$) compared to NFW. This is not
unexpected since these profiles are quite similar over a large region
\citep{VeraCiro2013MNRAS}. Also in the case of the combined evidence
the NFW and Einasto are hard to distinguish. Comparing the NFW or
Einasto profiles for individual galaxies to the cored models, one
cannot strongly rule out a particular model. For Fornax, Sculptor and
Carina, the $\gamma = 2$ models (where the transition speed is fast)
appear to be less likely, but this is not the case for Sextans. However, when we look
at the combined evidence, i.e. we explore whether all dwarfs are
embedded in the same halos, such $\gamma = 2$ models are clearly disfavored.

The results for Sculptor may be compared to those of
\citet{Breddels2012arXiv}.  In that paper, the authors found that the maximum
likelihood value for the central slope of the density profile corresponded to
a cored model. Since the evidence is the integral of the pdf, and not
directly related to the maximum likelihood (except for a Gaussian
distribution), we should not be surprised to find a slightly stronger
evidence for the NFW case here. In any case, the differences between
the models are minor as shown graphically in
Fig. \ref{fig:moments_bestfit}, and the evidence and the maximum
likelihood (marginally) favoring different models can be
attributed simply to not being able to distinguish amongst these.

\newcommand{\scaleFigMass}{0.55}
\begin{figure*}
\begin{tabular}{c @{}c@{} @{}c@{} @{}c@{} @{}c@{}}
 & \textbf{Fornax} & \textbf{Sculptor} & \textbf{Carina} & \textbf{Sextans}\\
& \includegraphics[scale=\scaleFigMass]{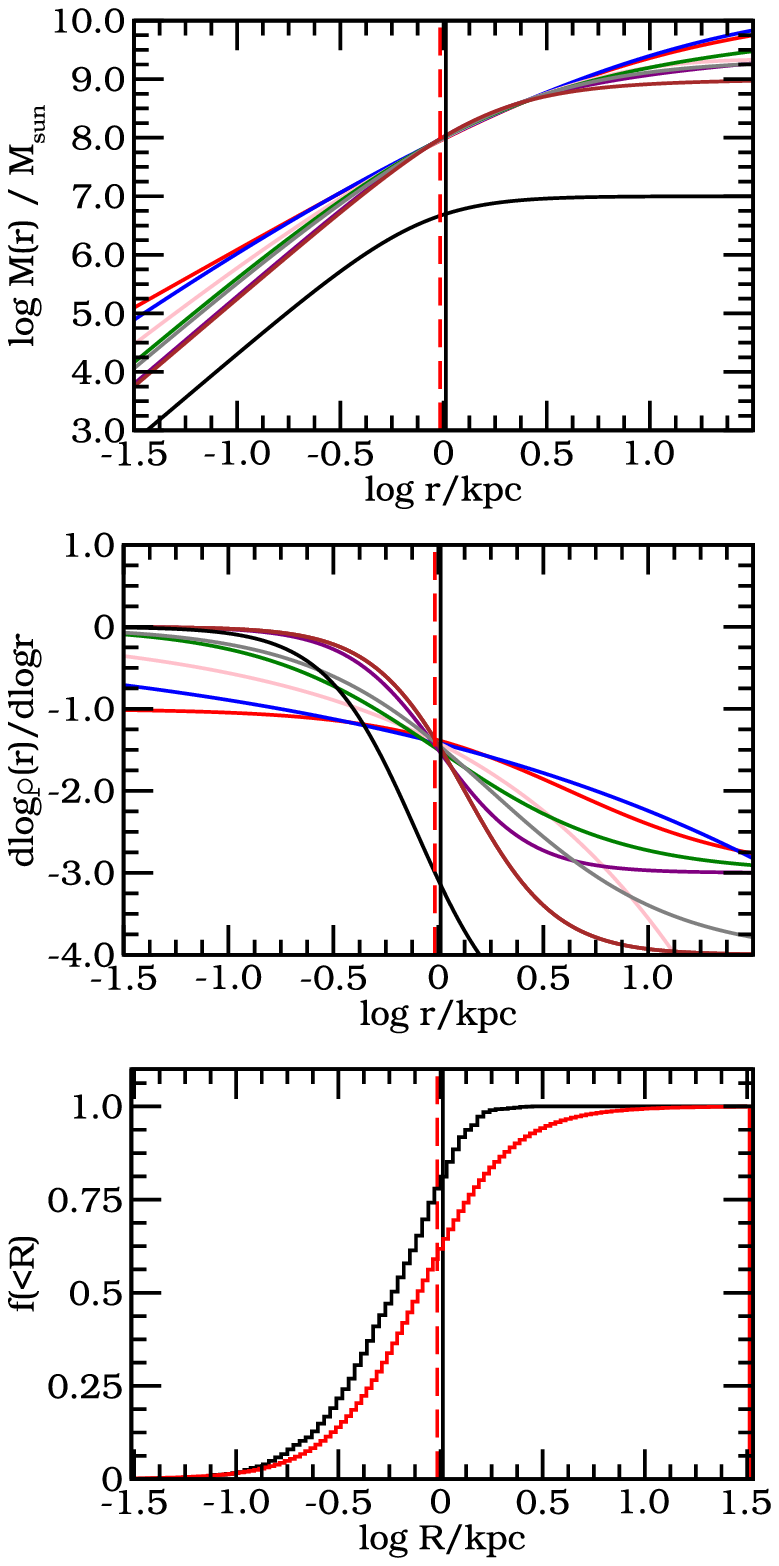} & \includegraphics[scale=\scaleFigMass]{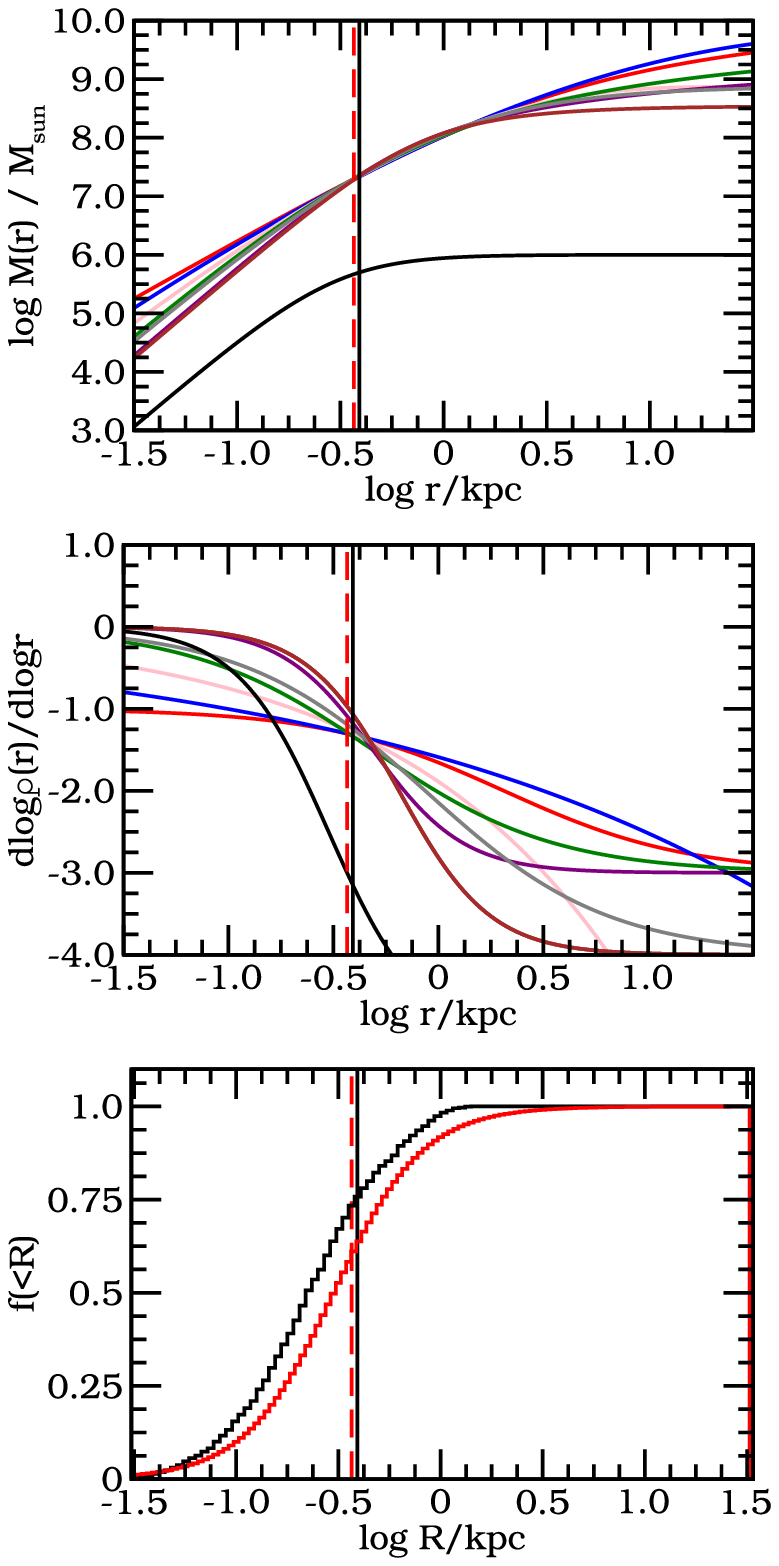} & \includegraphics[scale=\scaleFigMass]{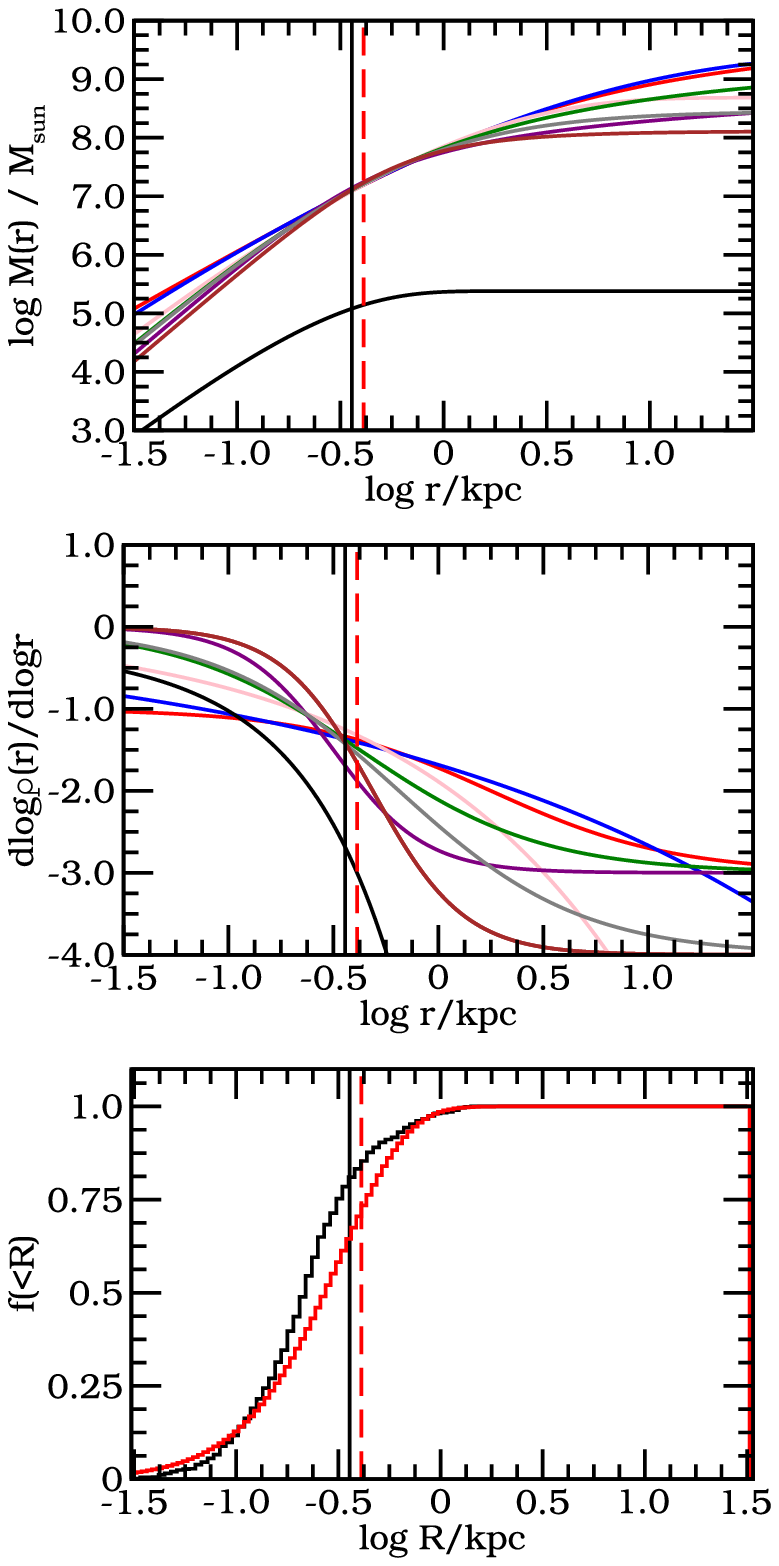} & \includegraphics[scale=\scaleFigMass]{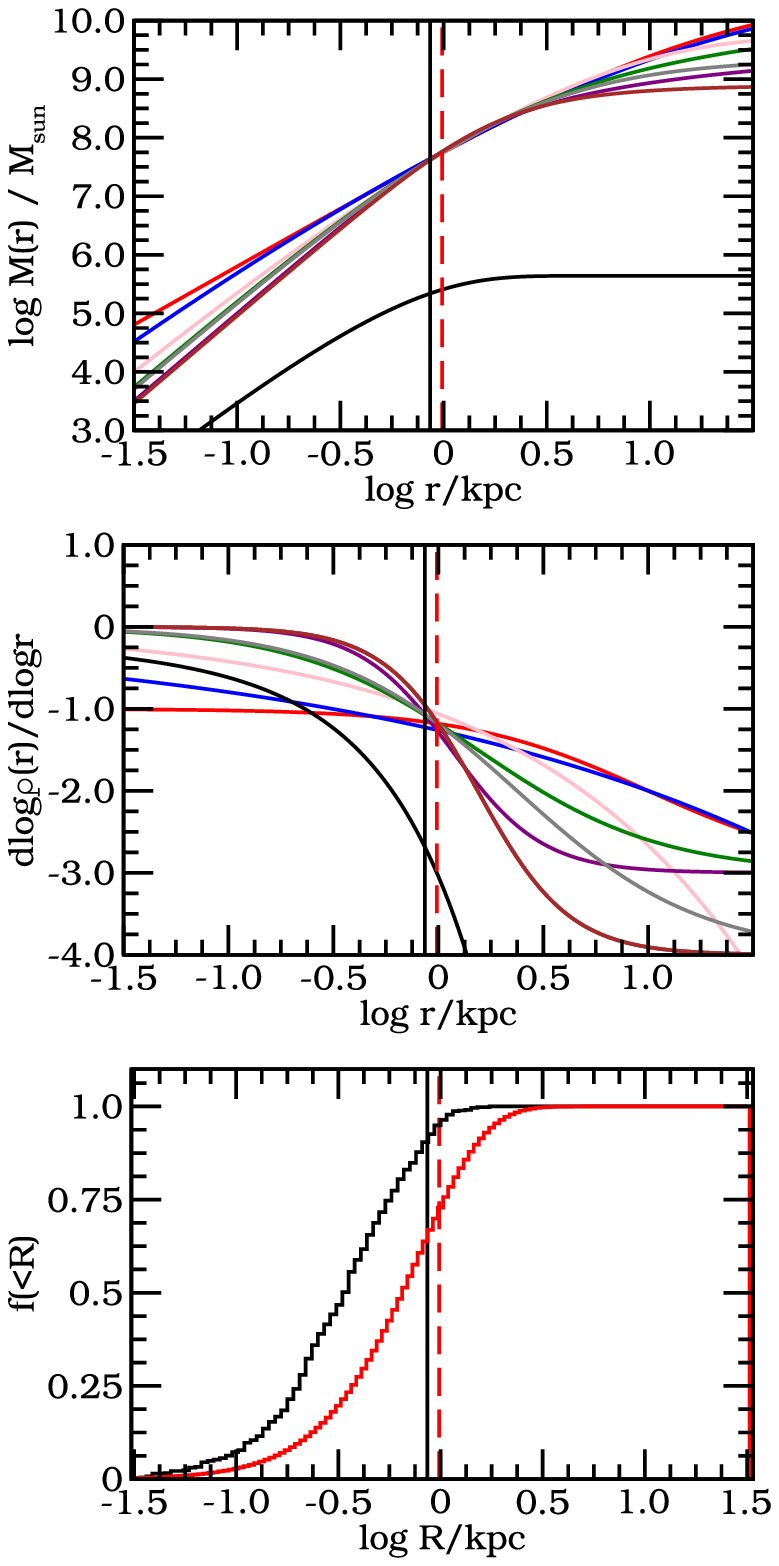}
\end{tabular}
\caption{{\bf Top row}: Enclosed mass as a function of radius for the different dark matter density profiles, with the stellar component in black.  {\bf Middle row}: Logarithmic density slope as a function or radius, where the black curve corresponds again to the stellar component. The red dashed line indicates $r_{-3}$, the radius at which the light profile has a logarithmic slope of $-3$, while the black line indicates $r_{1/2}$, the radius at which half of the stellar mass in enclosed (in 3d). {\bf Bottom row}: Cumulative density distribution of the (2d) radial distribution of the data (black), and the light (red) showing the kinematic data is sampled more concentrated towards the center. \label{fig:mass_profiles}}
\end{figure*}

\subsection{A robust slope measurement}
\label{sec:slope}

We now inspect in more detail the shape of the mass distributions found for the various
best fitting models. We are interested in exploring why the differences between the various
models as small as apparent in Fig.~\ref{fig:moments_bestfit}.

The top row of Fig.~\ref{fig:mass_profiles} shows the enclosed
dark matter mass for the best fit models (indicated by the solid dots
in Fig.~\ref{fig:pdfs}) for each galaxy separately. We use the same
color coding as in Fig.~\ref{fig:pdfs}, and also include the stellar
mass in black. The red-dashed vertical lines denote $r_{-3}$, the
radius at which the light density profile has a logarithmic slope of $-3$,
while the black line indicates $r_{1/2}$. This
remarkable figure shows that for each galaxy there is a region where
the mass distributions are truly almost indistinguishable from one
another. The different profiles, each characterized by its own
functional form, scale radius $r_s$ and mass $M_{\rm 1kpc}$, conspire to
produce a unique mass distribution. This region extends from slightly
below $r_{-3}$ to approximately the location of the outermost data
point (see bottom panel). Here $M(r) \propto r^x$, where $x$ ranges from $1.65$ for Fnx, to $1.9$ for Sextans.

In the middle row of Fig.~\ref{fig:mass_profiles} we plot the
logarithmic slope of the dark halo density distribution, where the
black line denotes the stellar density. Near the position where the
logslope of the stellar density is $-3$, all the best fit dark matter
density profiles seem to reach a similar logslope, although the value
of the slope varies from galaxy to galaxy.
The radius where the
logslopes coincide lies, as expected, inside the region where the mass distribution
is well determined, since both quantities are related through
derivatives. 

To illustrate the distribution of the kinematic sample with respect to
the light, we plot in the bottom row of  Fig. \ref{fig:mass_profiles}  the cumulative 2d radial
distribution of the kinematic data in black. The cumulative 2d radial
distribution for the light is plotted as the red histogram. All
kinematic datasets are more concentrated than the light, but no clear
trend is visible between the distribution of the kinematic sample with
respect to the light, and the exact location where the logslope of
most accurately determined.

The existence of a finite region where the mass is more accurately
determined has also been observed in the literature in works using MCMC in combination with Jeans
modeling. For example, it is visible in e.g. the
right panel of Fig 1. in \citet{Wolf2010MNRAS}, Fig. 18 in
\citet{Walker2012arXiv}, and Fig. 10 in \cite{Jardel2013ApJ} for
Draco, in the case of a non-parametric density distribution with
Schwarzschild models.

\newcommand{\scaleFigrThree}{0.6}
\begin{figure*}
\begin{tabular}{c @{}c@{} @{}c@{} @{}c@{} @{}c@{}}
 & \textbf{Fornax} & \textbf{Sculptor} & \textbf{Carina} & \textbf{Sextans}\\
& \includegraphics[scale=\scaleFigrThree]{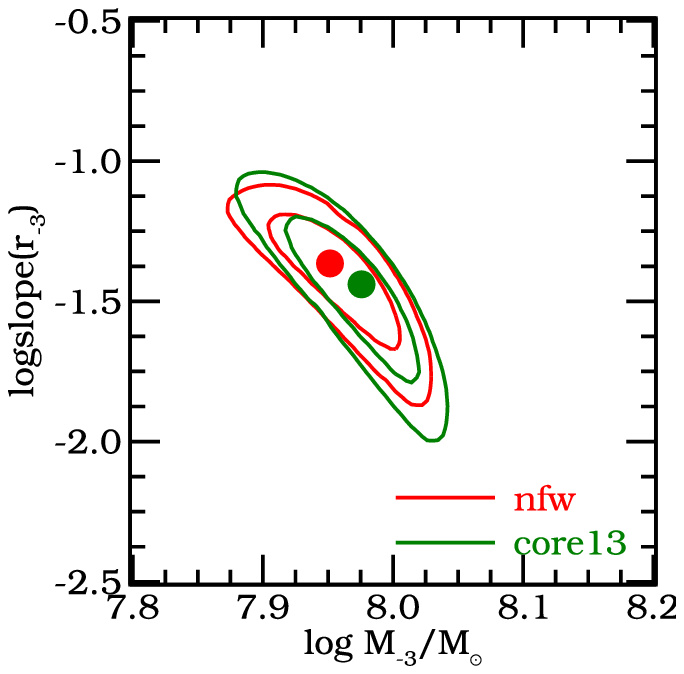} & \includegraphics[scale=\scaleFigrThree]{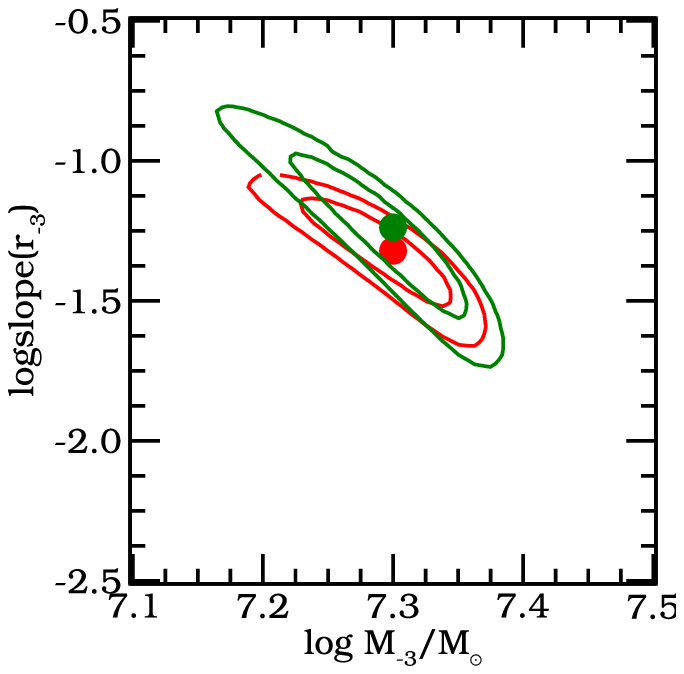} & \includegraphics[scale=\scaleFigrThree]{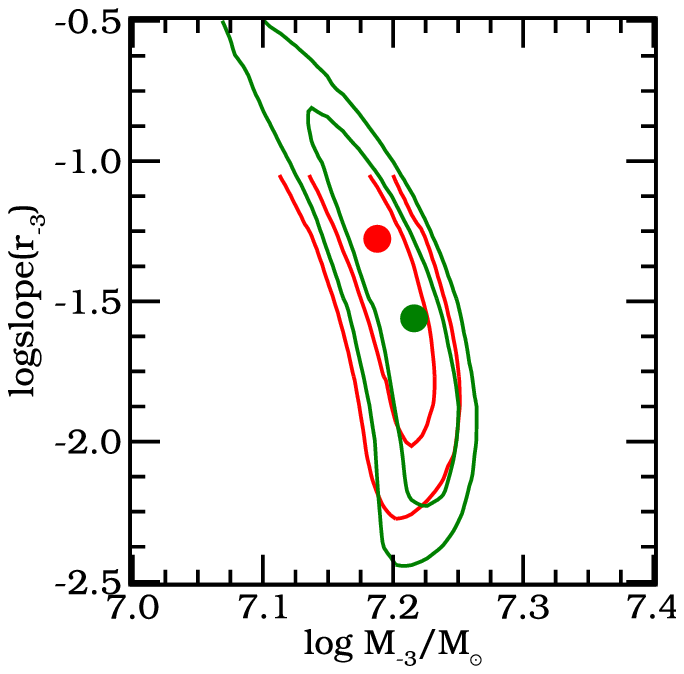} & \includegraphics[scale=\scaleFigrThree]{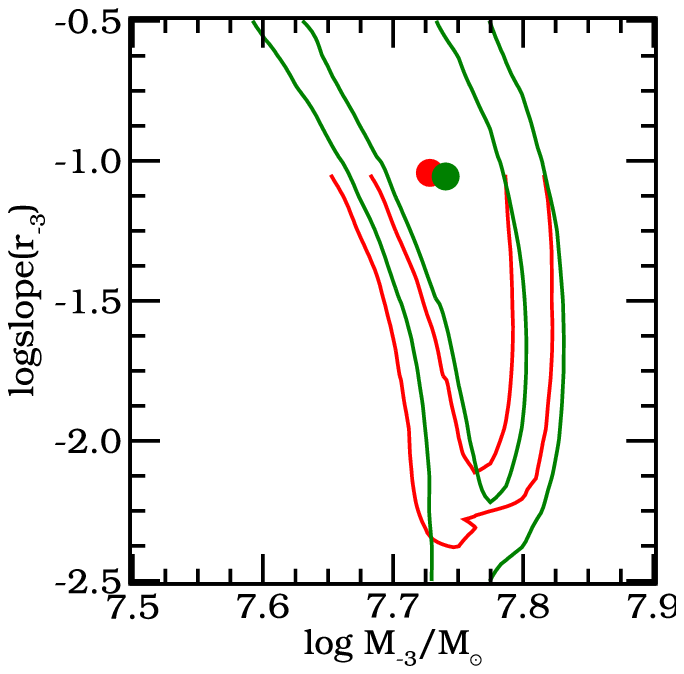}
\end{tabular}
\caption{Similar to Fig. \ref{fig:pdfs}, except now using $M_{-3}$ and $r_{-3}$ as parameters. Note that the contours for the NFW cannot go beyond $\kappa(r_{-3}) \ge -1$. \label{fig:pdfrThree}}
\end{figure*}

The analysis of \citet[][]{Wolf2010MNRAS} used the light weighted
average of the velocity dispersion to relate the radius at which the
logslope of the light is $-3$, or the half light radius, to the point
where the mass is accurately (being independent on the anisotropy) and
precisely (showing the least uncertainty) determined. Our findings go beyond this result. They
suggest that whatever dynamical model or
method is explored, there is a better set of parameters to describe
the mass distribution of dSph galaxies.  Let $r_{-3}$ be the radius at
which the logslope of the (3d) light distribution is $-3$. Since the
mass is accurately determined in this region, a natural parameter
would be $M_{-3}=M(r_{-3})$. And since also the logslope at this
radius is accurately determined, the next parameter should be
$\kappa_{-3} = \frac{\text{d}\log \rho}{\text{d}\log r} |_{r=r_{-3}}$.
For any general model, if the values of $\beta$ and $\gamma$ are fixed, this effectively makes $r_s$ a function of
$\kappa_{-3}$.

Fig.~\ref{fig:pdfrThree} shows the pdf for the $M_{-3}$ and
$\kappa_{-3}$ parameters for both the NFW and core13 models for each
galaxy, assuming a flat prior on these parameters in the domain shown
in this figure, except for the NFW profile which we limit to
$\kappa_{-3}=-1.05$, since for $\kappa_{-3} \ge -1$ the scale radius
is unphysical. As can be seen from the pdf while there is still
uncertainty associated to the logslope at this radius, the value is
nearly model independent and therefore we believe this value to be
accurate, especially for Fornax and Sculptor ($\kappa_{-3}=-1.4\pm0.15$ and $\kappa_{-3}=-1.3\pm0.12$ respectively for the NFW model).
Note also that some uncertainties might arise because the
kinematics are not sampled exactly according to the light.

These results also help us understand why we found that the scale radii of the best fitting
NFW profiles always to be larger than those of the cored models (see Fig.~\ref{fig:pdfs}).
For the NFW, we have
\begin{equation}
 \kappa(r) = \frac{{\rm d} \log \rho(r)}{{\rm d} \log r} = \frac{-2 r}{r + r_s} - 1,
\end{equation}
which can be easily solved for $r_s$:
\begin{equation}
 r_{s,\text{nfw}} = -r \frac{\kappa + 3}{\kappa + 1}.
\end{equation}
A similar solution can be found for the other parametric models, for instance the $\gamma\beta$ model gives:
\begin{equation}
 r_{s,\gamma\beta} = r \left( \frac{-\kappa}{\beta+\kappa} \right)^{-1/\gamma}.
\end{equation}
If we now require that the slopes are the same at $r_{-3}$ for the NFW and core13 models, we find
\begin{equation}
\frac{r_{s,\text{nfw}}}{r_{s,\text{core13}}} = \frac{\kappa}{\kappa+1},
\end{equation}
which is $> 1$ for $\kappa < -1$, explaining why the cored profiles
have smaller scale radii than the NFW profile, i.e. to get the same
logslope at the same location for the cored models, their scale radius
needs to be smaller than that of the NFW profile. A similar result holds for
the other cored models.

\section{Conclusions}
\label{sec:end}

In this paper we have presented a comparison of dynamical models using
different dark matter profiles for four dwarf spheroidal galaxies in
the Local Group, namely Fornax, Sculptor, Carina and Sextans. The
model comparison was done using Bayesian evidence. We have found that
no particular model is significantly preferred, and that all four
dwarf spheroidals are compatible with either NFW/Einasto or any of the
explored cored profiles. Only Sextans shows a slight preference for
cored models, but not with high odds. Nonetheless, we find that it is
very unlikely that all four dwarf spheroidals are each embedded in a
cored dark matter halo of the form $\rho_{DM} \propto 1/(1 +
r^2)^{\beta/2}$, with $\beta = 3, 4$.

Our best fit models however, conspire to produce the same mass
distribution over a relatively large range in radii, from $r_{-3}$
up to the last measured data point (which is often close to the nominal
tidal radius obtained from fitting the light profile). This $M(r) \sim
r^{x}$, with $x = 1.65-1.9$, is similar to that suggested by \citet{Walker2009} albeit with a slightly
steeper exponent. Another (related) quantity that is robustly
determined and independent of the assumed dark matter density profile,
is the logslope of the density distribution at $r_{-3}$. We find for
the dwarfs in our sample, that this slope ranges from $\kappa_{-3} \sim
-1.4$ at $r_{-3} = 0.96$ kpc for Fornax, to $\kappa_{-3} \sim -1.1$ at $r_{-3} =
0.98$ kpc for Sextans.

These findings can be seen as an extension of the results of
\citet{Wolf2010MNRAS}, who showed that the mass at $r_{-3}$ can be
determined very accurately in a model independent fashion. These
authors demonstrated that this result might be understood from the
Jeans equation. Although we do not have yet a solid mathematical
explanation for our new findings, we suspect that this might be
obtained using the virial theorem, which is effectively another, yet
independent moment of the collisionless Boltzmann equation.

In the near future, we will apply Schwarzschild modeling to the same
data but instead of the moments, we will use the discrete
individual measurements directly. This approach should allows us to
get the most out of the data, since no information is lost. When
binning, one loses spatial resolution, but also the higher moments of
the line-of-sight velocity distribution are not included in the
fitting procedure because of their large and asymmetric errors.
Furthermore, the use of the full line-of-sight velocity distribution
should improve the precision of the anisotropy profile, which may be
an interesting quantity to discriminate formation scenarios.

This moments-to-discrete modeling step must be carried out before
deciding if and how much more data is needed to discriminate among
various dark matter density profiles. Nonetheless, we have learned
here that the functional form of the mass distribution may be
determined over a large distance range, even when only a few hundred
velocity measurements are available (as in the case of
Sextans). However, the uncertainty on the value of the exact slope of
the density profile at e.g. $r_{-3}$ is driven by the sample size.

An obvious next step is to establish if the subhalos extracted from
cosmological simulations have the right characteristics to host the
dSph of the Milky Way, now that not only the mass, but also its
functional form (1st and 2nd derivatives), of their dark halos have
been determined reliably.

\section*{Acknowledgments}
We are grateful to Giuseppina Battaglia, Glenn van de Ven and Remco van den Bosch for
discussions that led to the work presented here. We acknowledge financial support from
NOVA (the Netherlands Research School for Astronomy), and
European Research Council under ERC-StG grant GALACTICA-24027.

\bibliography{schwdsphs}

\end{document}